\documentclass[5p]{elsarticle}
\usepackage{url}
\usepackage{bm}		
\usepackage{graphicx}	
\usepackage{dcolumn}	
\usepackage{fancybox}	
\usepackage{longtable}	
\usepackage{fontenc}
\usepackage{amsmath}
\usepackage{amssymb}

\begin{document}


\title{The physics of the relativistic counter-streaming instability that drives mass inflation inside black holes}

\author[jila]{Andrew J. S. Hamilton}
\ead{Andrew.Hamilton@colorado.edu}
\address[jila]{JILA and
Dept.\ Astrophysical \& Planetary Sciences,
Box 440, U. Colorado, Boulder, CO 80309, USA}
\author[porto]{Pedro P. Avelino}
\ead{ppavelin@fc.up.pt}
\address[porto]{Departamento de F\'isica da Faculdade de Ci\^encias da Universidade do Porto and Centro de F\'isica do Porto, Rua do Campo Alegre 687, 4169-007 Porto, Portugal}

\newcommand{\dd}{d}
\newcommand{\DD}{D}
\newcommand{\im}{i} 
\newcommand{\ee}{e} 
\newcommand{\perpperp}{\perp\!\!\perp}
\newcommand{\nn}{\nonumber\\}
\newcommand{\simpropto}{
\raisebox{-0.6ex}[1.5ex][0ex]{$
    \begin{array}[b]{@{}c@{\;}} \propto \\
    [-1.75ex] \sim \end{array}
  $}
}
\newcommand{\Msun}{\textrm{M}_\odot}
\newcommand{\vel}{\textsl{v}}
\newcommand{\inn}{\textrm{in}}
\newcommand{\out}{\textrm{ou}}
\newcommand{\coll}{\textrm{coll}}
\newcommand{\unit}[1]{\, \rm{#1}}

\newcommand{\Mbh}{M_\bullet}
\newcommand{\Mbhdot}{\dot{M}_\bullet}
\newcommand{\Qbh}{Q_\bullet}
\newcommand{\Planck}{\textrm{p}}
\newcommand{\TUnruh}{T_\textrm{U}}

\newcommand{\bg}{\bm{g}}
\newcommand{\bp}{\bm{p}}
\newcommand{\bx}{\bm{x}}
\newcommand{\bgamma}{\bm{\gamma}}
\newcommand{\bpartial}{\bm{\partial}}

\hyphenpenalty=3000

\newcommand{\betadiagramfig}{
    \begin{figure}[t!]
    \begin{center}
    \leavevmode
    \includegraphics[scale=.7]{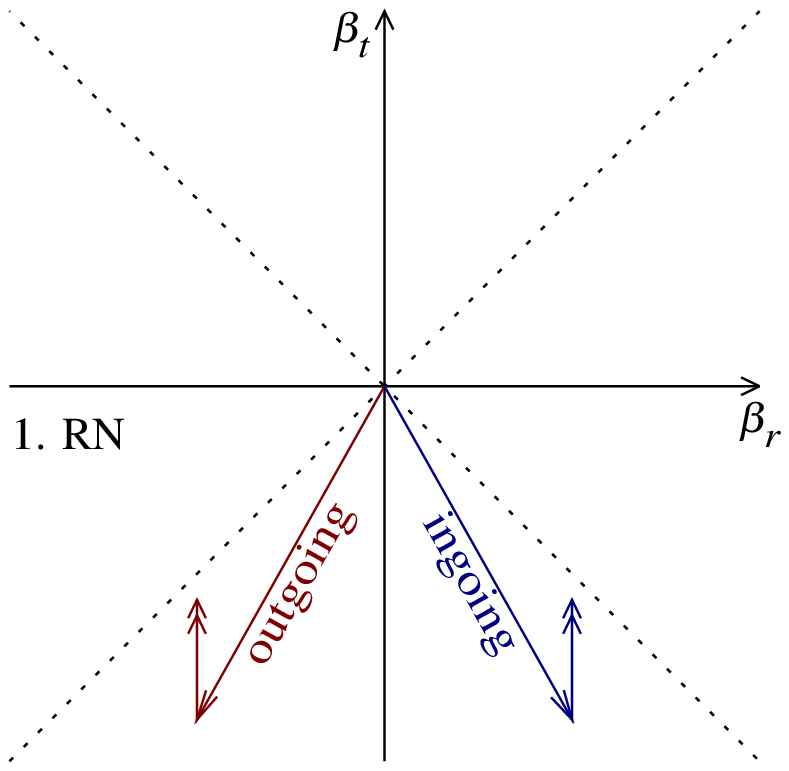}
    \vskip2ex
    \includegraphics[scale=.7]{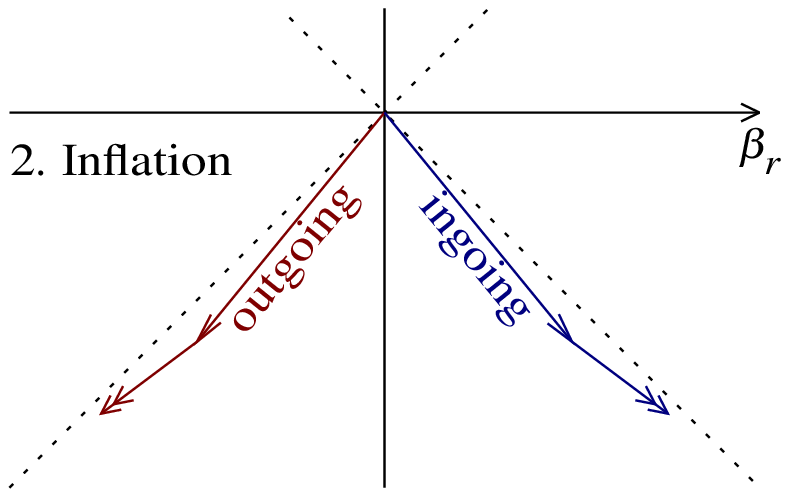}
    \vskip2ex
    \includegraphics[scale=.7]{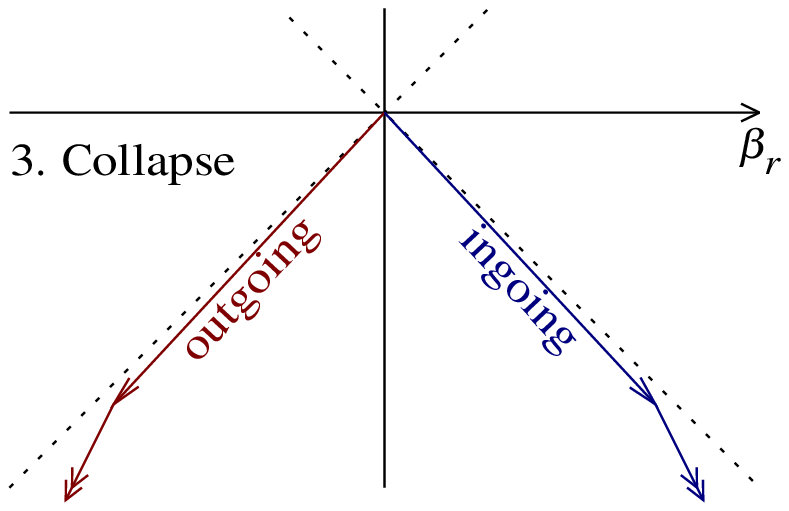}
    \caption[1]{
    \label{betadiagram}
Spacetime diagrams
of the tetrad-frame 4-vector $\beta_m$, equation~(\protect\ref{betam}),
illustrating qualitatively the three successive phases of mass inflation:
1.~(top) the Reissner-Nordstr\"om phase,
where inflation ignites;
2.~(middle) the inflationary phase itself;
and
3.~(bottom) the collapse phase,
where inflation comes to an end.
In each diagram,
the arrowed lines
illustrate two representative examples of the 4-vector
$\{ \beta_t , \beta_r \}$,
one ingoing and one outgoing,
while the double-arrowed lines illustrate the rate of change
of these 4-vectors
implied by Einstein's equations~(\protect\ref{dtbspherical}).
Inside the horizon of a black hole,
all locally inertial frames necessarily fall inward,
so the radial velocity $\beta_t \equiv \partial_t r$
is always negative.
A locally inertial frame
is ingoing or outgoing depending on whether
the proper radial gradient
$\beta_r \equiv \partial_r r$
measured in that frame is positive or negative.
    }
    \end{center}
    \end{figure}
}

\newcommand{\budiagramfig}{
    \begin{figure}[t!]
    \begin{center}
    \leavevmode
    \includegraphics[scale=.7]{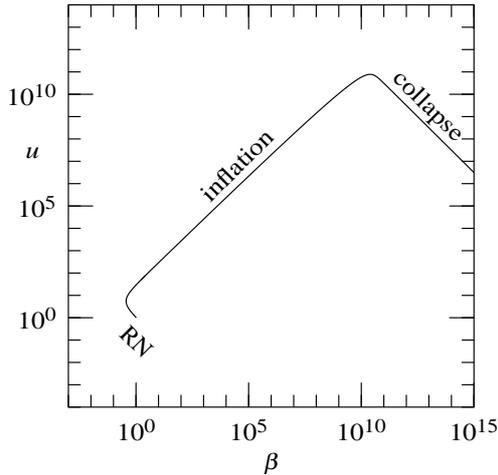}
    \caption[1]{
    \label{budiagram}
The variation of $\beta$ and $u$
predicted by the simple model of two equal,
pressureless, ingoing and outgoing streams.
The parameters are $\lambda = 3$ and $\mu = 0.1$,
and $\beta$ and $u$ are both initially $1$.
    }
    \end{center}
    \end{figure}
}

\newcommand{\bufig}{
    \begin{figure}[t!]
    \begin{center}
    \leavevmode
    \includegraphics[scale=.7]{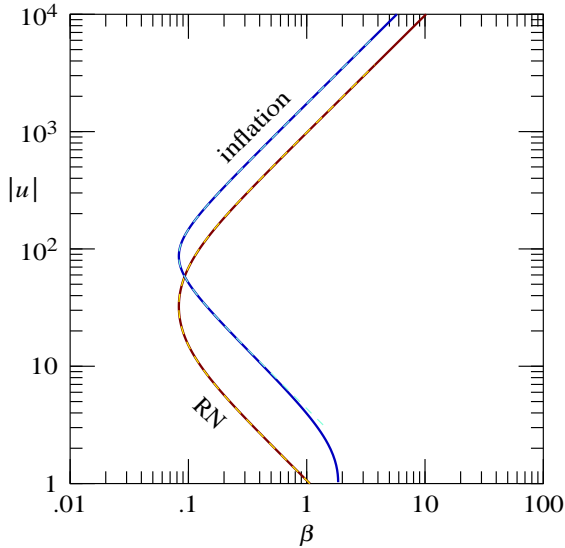}
    \caption[1]{
    \label{bu}
Absolute values $\lvert u \rvert$ of velocities of
outgoing baryonic
(light blue)
and ingoing dark matter
(dark brown)
streams
as a function of $\beta$
in a self-similar solution with
charged, almost pressureless baryons
and neutral, pressureless dark matter.
The superposed dashed lines
are the analytic approximation given by equation~(\protect\ref{betaami}).
The parameters of the model and of the fit are listed in Table~\ref{par}.
    }
    \end{center}
    \end{figure}
}

\newcommand{\bucollapsefig}{
    \begin{figure}[t!]
    \begin{center}
    \leavevmode
    \includegraphics[scale=.7]{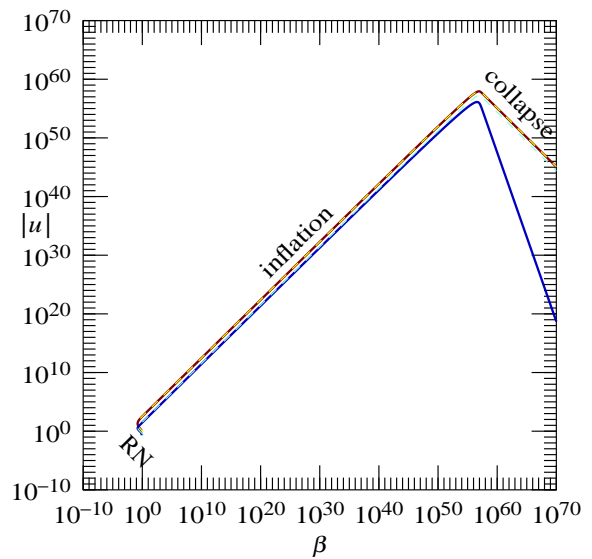}
    \caption[1]{
    \label{bucollapse}
Similar to Figure~\protect\ref{bu},
illustrating all three phases of evolution
in a self-similar solution similar to that in Figure~\protect\ref{bu},
but in which the baryons,
instead of being almost pressureless,
have a relativistic equation of state $w \equiv p_b / \rho_b = 0.32$.
The superposed dashed lines
are the analytic approximation~(\protect\ref{betaami})
for the Reissner-Nordstr\"om+inflation phases,
and the analytic approximation~(\protect\ref{ubacollapse})
for the inflationary+collapse phases.
    }
    \end{center}
    \end{figure}
}

\newcommand{\buunbalancedfig}{
    \begin{figure}[t!]
    \begin{center}
    \leavevmode
    \includegraphics[scale=.7]{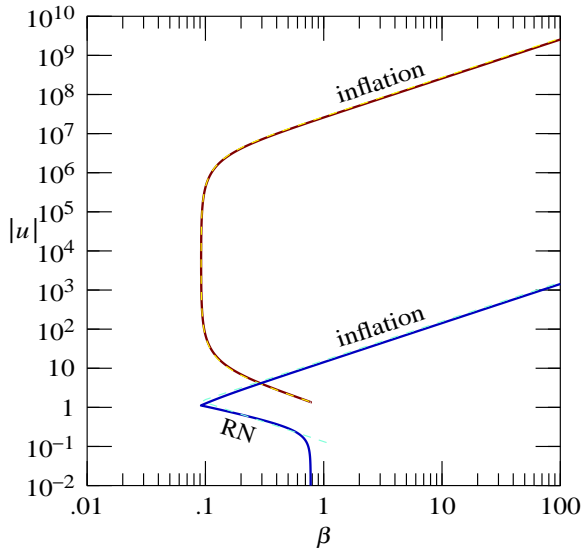}
    \caption[1]{
    \label{buunbalanced}
Similar to Figure~\protect\ref{bu},
illustrating a self-similar solution in which one stream,
the ingoing dark matter stream
(upper dark brown line),
is much smaller than the other,
the outgoing baryonic stream
(lower light blue line).
The superposed dashed lines
are the analytic approximation~(\protect\ref{betaami}).
    }
    \end{center}
    \end{figure}
}

\newcommand{\darkmatteraccretionMfig}{
    \begin{figure}[t]
    \begin{center}
    \leavevmode
    \includegraphics[scale=.7]{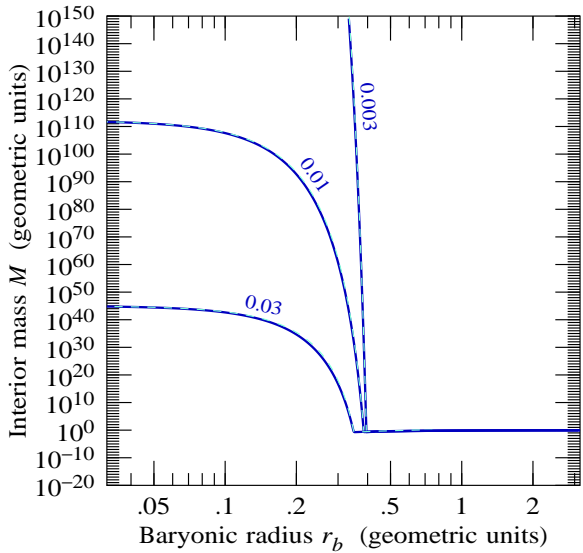}
    \caption[1]{
    \label{darkmatteraccretionM}
Interior mass $M$, equation~(\protect\ref{M}),
as a function of baryonic radius $r_b$
for models with dark-matter-to-baryon ratio
$\rho_d/\rho_b = 0.1$
at the sonic point,
and three different accretion rates,
$\Mbhdot = 0.003$, $0.01$, $0.03$.
The smaller the accretion rate,
the faster the interior mass $M$ inflates.
The superposed dashed lines
are the analytic approximation~(\protect\ref{rami})
for the Reissner-Nordstr\"om+inflation phases,
and the analytic approximation~(\protect\ref{racollapse})
for the inflationary+collapse phases.
If the interior mass were plotted against the dark matter radius $r_d$
instead of the baryonic radius $r_b$,
then the lines would be different, though similar.
The parameters of the models and of the fits are listed in Table~\ref{par}.
    }
    \end{center}
    \end{figure}
}

\newcommand{\darkmatteraccretionfig}{
    \begin{figure}[t]
    \begin{center}
    \leavevmode
    \includegraphics[scale=.7]{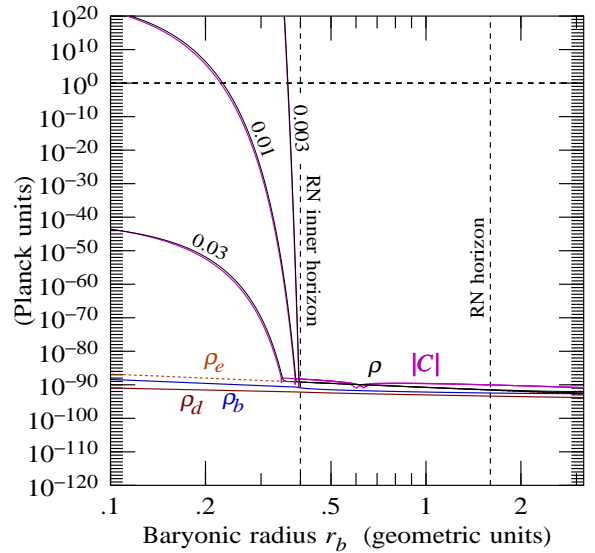}
    \caption[1]{
    \label{darkmatteraccretion}
Center-of-mass energy density $\rho$ (thin black lines)
and Weyl curvature scalar $\lvert C \rvert$
(purple lines almost coincident with the thin black lines)
for the models shown in Figure~\protect\ref{darkmatteraccretionM},
labeled with their mass accretion rates
$\Mbhdot = 0.003$, $0.01$, $0.03$.
The center-of-mass energy density
and Weyl curvature
inflate like the interior mass $M$.
Also shown are
the individual proper densities $\rho_b$ of baryons,
$\rho_d$ of dark matter, and $\rho_e$ of electromagnetic energy
for the model with
$\Mbhdot = 0.01$
(to avoid confusion, only this case is plotted;
the other models are similar).
During inflation,
almost all the center-of-mass energy $\rho$
is in the streaming energy:
the proper densities of individual components remain small.
Dashed vertical lines mark where the outer and inner horizons
would be in the Reissner-Nordstr\"om geometry with the same
charge-to-mass $\Qbh/\Mbh = 0.8$;
mass inflation destroys the inner horizon.
The density scale is in Planck units
for a black hole of mass $\Mbh = 4 \times 10^6 \unit{\Msun}$.
    }
    \end{center}
    \end{figure}
}

\newcommand{\eqstatefig}{
    \begin{figure}[tp!]
    \begin{center}
    \leavevmode
    \includegraphics[scale=.7]{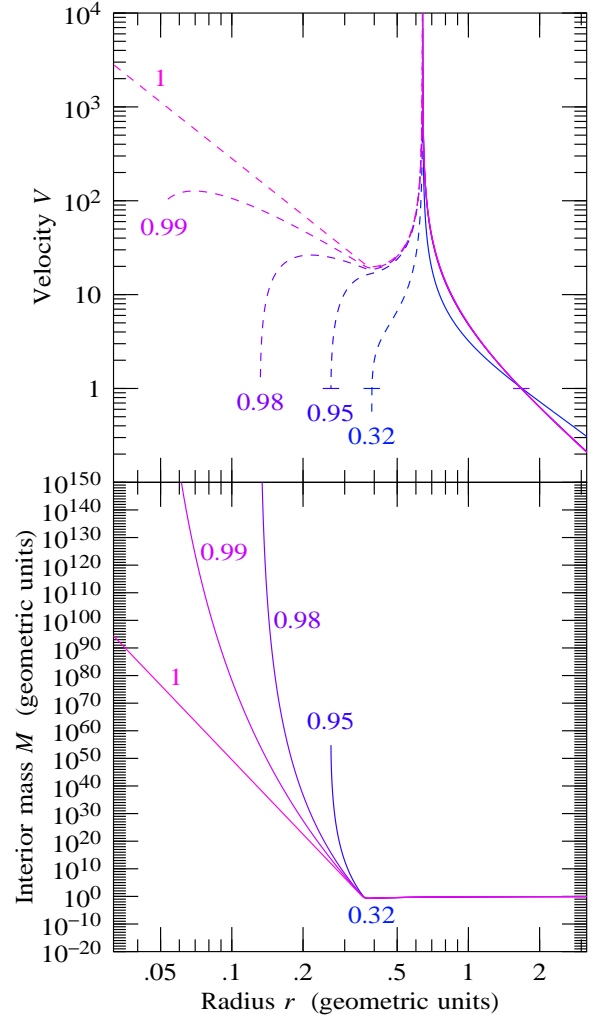}
    \caption[1]{
    \label{eqstate}
Proper velocity $V$ (top)
of
the similarity frame
relative to
the fluid,
and interior mass $M$ (bottom),
as a function of radius $r$
in charged single fluid models
with various equations of state $w \equiv p / \rho$.
Curves are labeled with $w$.
Mass inflation begins but then stalls
in models with sound speed near but less than the speed of light,
$w < 1$.
Only when the sound speed equals the speed of light,
$w = 1$,
does mass inflation continue.
Velocity lines are dashed where $V$ is negative.
Short horizontal bars
mark the positions of outer and inner horizons,
where $\lvert V \rvert = 1$.
The black hole charge-to-mass
$\Mbh/\Qbh = 0.8$
and
accretion rate $\Mbhdot = 0.01$
are the same in all cases.
    }
    \end{center}
    \end{figure}
}

\newcommand{\scalarfieldaccretionfig}{
    \begin{figure}[t]
    \begin{center}
    \leavevmode
    \includegraphics[scale=.7]{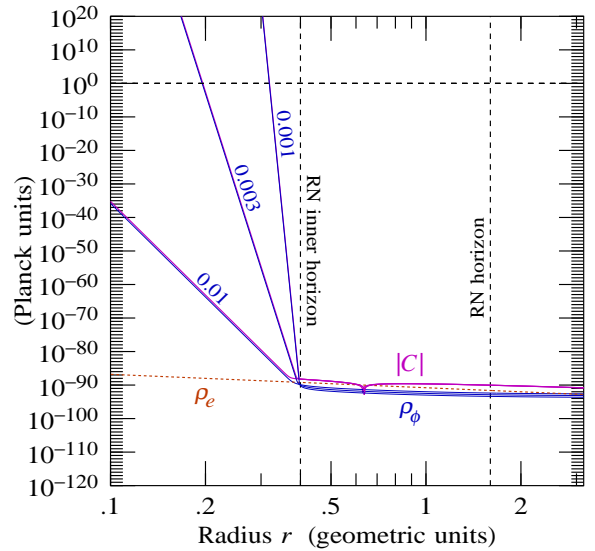}
    \caption[1]{
    \label{scalarfieldaccretion}
Fluid energy density $\rho_\phi$ (thin blue lines)
and Weyl curvature scalar $\lvert C \rvert$
(purple lines almost coincident with the thin blue lines)
inside a black hole accreting
a perfect fluid with an ultra-hard equation of state,
$w \equiv p_\phi / \rho_\phi = 1$.
Lines are labeled with their mass accretion rates
$\Mbhdot = 0.001$, $0.003$, and $0.01$.
Also shown is the density
$\rho_e$ of electromagnetic energy,
which is the same for all the models.
The graph illustrates that
the smaller the accretion rate,
the faster the density and curvature inflate.
Dashed vertical lines mark where the outer and inner horizons
would be in the Reissner-Nordstr\"om geometry with the same
charge-to-mass $\Qbh/\Mbh = 0.8$.
The parameters of the models are listed in Table~\ref{par}.
The density scale is in Planck units
for a black hole of mass $\Mbh = 4 \times 10^6 \unit{\Msun}$.
    }
    \end{center}
    \end{figure}
}

\newcommand{\scalarmdotfig}{
    \begin{figure}[tb]
    \begin{center}
    \leavevmode
    \includegraphics[scale=.7]{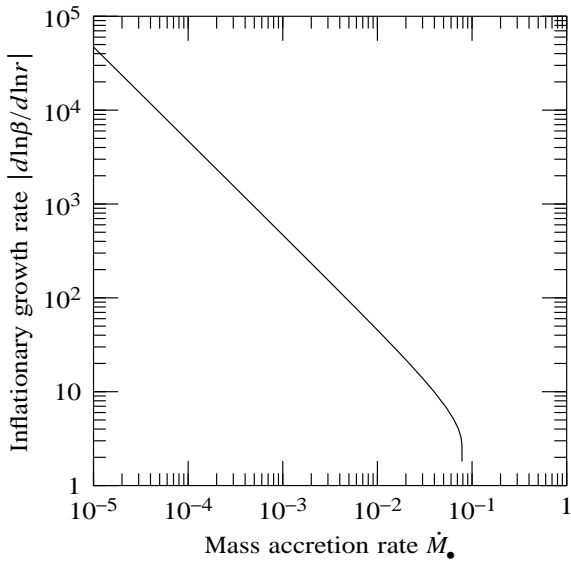}
    \caption[1]{
    \label{scalarmdot}
Inflationary growth rate
$| \dd \ln \beta / \dd \ln r |$
as a function of mass accretion rate
$\Mbhdot$
in self-similar models
in which
an ultra-hard fluid
accretes on to a black hole with
charge-to-mass $\Qbh/\Mbh = 0.8$.
The inflationary growth rate is inversely proportional to the
mass accretion rate,
equation~(\protect\ref{scalardlnbdlnr}),
except at the largest accretion rates.
The curve terminates at
a maximum accretion rate of
$\Mbhdot = 0.0784$.
    }
    \end{center}
    \end{figure}
}

\newcommand{\bhcolliderfig}{
    \begin{figure}[t!]
    \begin{center}
    \leavevmode
    \includegraphics[scale=.7]{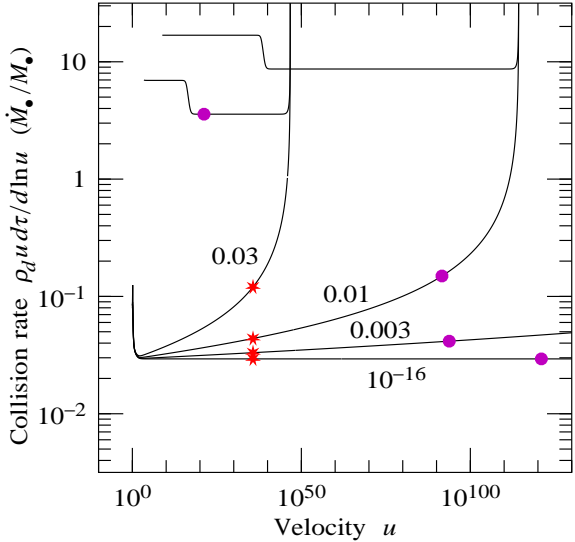}
    \caption[1]{
    \label{bhcollider}
Collision rate of the black hole particle accelerator
per $\ee$-fold of velocity $u$ (meaning $\gamma v$),
expressed in units of the inverse black hole accretion time
$\Mbhdot / \Mbh$.
The models illustrated are the same as those in
Figure~\ref{darkmatteraccretion}.
The curves are labeled with
their mass accretion rates:
$\Mbhdot = 0.03$, $0.01$, $0.003$, and $10^{-16}$.
Stars mark where the center-of-mass energy
of colliding baryons and dark matter particles exceeds the Planck energy,
while disks show where
the Weyl curvature scalar $C$ exceeds the Planck scale.
    }
    \end{center}
    \end{figure}
}

\newcommand{\phasetab}{
    \begin{table}[t!]
    \begin{center}
    \caption{
    \label{phase}
Phases of evolution.
    }
    \begin{tabular}{llc}
\hline
Phase & Gravity & Dominant term \\
\hline
Reissner-Nordstr\"om & charge & $\lambda$ \\
Inflation & streaming & $\mu u^2$ \\
Collapse & mass & $\beta^2$ \\
\hline
    \end{tabular}
    \end{center}
    \end{table}
}

\newcommand{\partab}{
    \begin{table*}[t!]
    \begin{center}
    \caption{
    \label{par}
Model and fit parameters.
    }
    \begin{tabular}{lccccccccc}
\hline
 & \multicolumn{4}{c}{\hrulefill\raisebox{-.5ex}{ Model parameters }\hrulefill} & \multicolumn{5}{c}{\hrulefill\raisebox{-.5ex}{ Fit parameters }\hrulefill} \\
Model & $\Qbh/\Mbh$ & $\Mbhdot$ & $w$ & $\rho_d/\rho_b$ & $C_b$ & $C_d$ & $\lambda$ & $\mu_b$ & $\mu_d$ \\
\hline
Fig.~\protect\ref{bu} & $0.8$ & $0.01$ & $10^{-6}$ & $0.001$ & $4.35$ & $-1.06$ & $14.6$ & $1.93 \times 10^{-3}$ & $1.39 \times 10^{-2}$ \\
Fig.~\protect\ref{buunbalanced} & $0.8$ & $0.01$ & $0.32$ & $10^{-6}$ & $0.133$ & $-1.05$ & $3.18$ & $1.51$ & $1.11 \times 10^{-7}$ \\
Figs.~\protect\ref{darkmatteraccretionM}, \protect\ref{darkmatteraccretion} & $0.8$ & $0.00316$ & $0.32$ & $0.1$ & $0.38$ & $-1.02$ & $3.07$ & $7.37 \times 10^{-2}$ & $2.80 \times 10^{-3}$ \\
Figs.~\protect\ref{bucollapse}, \protect\ref{darkmatteraccretionM}, \protect\ref{darkmatteraccretion} & $0.8$ & $0.01$ & $0.32$ & $0.1$ & $0.394$ & $-1.05$ & $3.25$ & $0.238$ & $8.88 \times 10^{-3}$ \\
Figs.~\protect\ref{darkmatteraccretionM}, \protect\ref{darkmatteraccretion} & $0.8$ & $0.0316$ & $0.32$ & $0.1$ & $0.431$ & $-1.18$ & $4.08$ & $0.767$ & $2.86 \times 10^{-2}$ \\
Fig.~\protect\ref{eqstate} & $0.8$ & $0.01$ & various & $0$ & $-$ & $-$ & $-$ & $-$ & $-$ \\
Fig.~\protect\ref{scalarfieldaccretion} & $0.8$ & various & $1$ & $0$ & $-$ & $-$ & $-$ & $-$ & $-$ \\
\hline
    \end{tabular}
    \end{center}
    \end{table*}
}

\begin{abstract}
If you fall into a real astronomical black hole
(choosing a supermassive black hole,
to make sure that the tidal forces don't get you first),
then you will probably meet your fate not at a central singularity,
but rather in the
exponentially growing, relativistic counter-streaming instability
at the inner horizon
first pointed out by Poisson \& Israel (1990),
who called it mass inflation.
The chief purpose of this paper is to present a clear exposition
of the physical cause and consequence of inflation
in spherical, charged black holes.
Inflation acts like a particle accelerator
in that it accelerates cold ingoing and outgoing streams
through each other to prodigiously high energies.
Inflation feeds on itself:
the acceleration is powered by the gravity produced by the streaming energy.
The paper:
(1)
uses physical arguments to
develop simple approximations that follow the evolution of inflation
from ignition, through inflation itself, to collapse;
(2)
confirms that the simple approximations capture accurately
the results of fully nonlinear
one- and two-fluid self-similar models;
(3)
demonstrates that, counter-intuitively,
the smaller the accretion rate,
the more rapidly inflation exponentiates;
(4)
shows that in single perfect-fluid models,
inflation occurs only if the sound speed equals the speed of light,
supporting the physical idea that inflation in single fluids
is driven by relativistic counter-streaming of waves;
(5)
shows that what happens during inflation up to the Planck curvature
depends not on the distant past or future,
but rather on events happening only a few hundred black hole crossing times
into the past or future;
(6)
shows that, if quantum gravity does not intervene,
then the generic end result of inflation is not
a general relativistic null singularity,
but rather a spacelike singularity at zero radius.
\end{abstract}

\begin{keyword}
\PACS 04.20.-q
\end{keyword}

\date{\today}

\maketitle

\tableofcontents

\section{Introduction}

It has been known for decades that something strange and dramatic must happen
near the inner horizon of a realistic black hole.
Penrose (1968) \cite{Penrose:1968} page 222
first pointed out that a person passing through the Cauchy horizon
(the outgoing inner horizon)
of an empty charged
(Reissner-Nordstr\"om)
or rotating
(Kerr-Newman)
black hole
will see the outside universe infinitely blueshifted.
Empty here means empty of charge, matter, or radiation
(but not of a static electric field)
except at the singularity.
Subsequent perturbation theory investigations
of the Reissner-Nordstr\"om geometry,
starting with
\cite{Simpson:1973}
and culminating with
\cite{Chandrasekhar:1982},
showed that perturbations from the outside world
would amplify to an infinite flux of energy on the Cauchy horizon,
a result that was interpreted as
indicating that the Reissner-Nordstr\"om geometry was unstable.
An early attempt to go beyond the linear regime is
\cite{Hiscock:1981}.


The full nonlinear nature of the instability near the inner horizon
was eventually clarified
in a seminal paper by Poisson \& Israel (1990)
\cite{Poisson:1989zz,Poisson:1990eh}.
Poisson \& Israel argued that,
if ingoing and outgoing streams are simultaneously present
just above the inner horizon of a spherical charged black hole,
then relativistic counter-streaming between the ingoing and outgoing streams
will lead to an exponentially growing instability which they dubbed
``mass inflation.''
During mass inflation,
the interior, or Misner-Sharpe \cite{Misner:1964}, mass,
a gauge-invariant scalar quantity,
exponentiates to huge values.
Other gauge-invariant measures, such as
the proper density, the proper radial pressure, and the Weyl scalar,
exponentiate along with the interior mass.
The phenomenon of mass inflation has been confirmed
analytically and numerically
in many papers
\cite{Ori:1991,Bonanno:1994qh,Brady:1995ni,Burko:1997fc,Burko:1997zy,Burko:2002fv,Dafermos:2004jp,Hansen:2005am,Hamilton:2004aw}.

Real black holes probably have very little charge,
thanks to the huge charge-to-mass ratio of individual protons and electrons,
$e / m_p \approx 10^{18}$,
where $e$ is the dimensionless charge of the proton or electron,
the square root of the fine-structure constant,
and $m_p$ is the proton mass in units of the Planck mass.
A charged black hole would quickly attract opposite charge and almost
neutralize, though the black hole would probably retain a small residual
positive charge because protons are more massive than electrons,
therefore more easily able to overcome a Coulomb barrier.
Whereas real black holes have little charge,
they probably do rotate (and rotate rapidly).
The interior structure of a charged black hole
resembles that of a rotating black hole
in that the negative pressure (tension) of the electric field
produces a gravitational repulsion
analogous to that produced by the centrifugal force in a rotating black hole.
As a result, charged spherical black holes have inner horizons
like rotating black holes.
It is thus common in modeling black hole interiors
to take charge as a surrogate for angular momentum,
e.g.~\cite{Poisson:1990eh,Dafermos:2004jp,Ori:2006fv},
and the present paper follows this route.
Further work on the interiors of rotating black holes is desirable,
but the problem is hard, and investigations remain few
\cite{Ori:1992,Ori:2001pc,Brady:1995un,Brady:1998ht,Hamilton:2009hu}
(see also the review \cite{Brady:1999}).

Much of the literature on mass inflation,
starting with \cite{Poisson:1990eh},
has focused on the situation where a black hole collapses,
and then remains isolated.
In this case,
perturbations scattering off the curvature of the collapsing black hole
are expected to lead to a decaying Price tail
\cite{Price:1972,Dafermos:2003yw}
of outgoing gravitational radiation,
as originally proposed by \cite{Poisson:1990eh}.
This outgoing radiation has usually been modeled
with a massless scalar field
\cite{Christodoulou:1986a,Christodoulou:1986b,Christodoulou:1987a,Christodoulou:1987b,Goldwirth:1987,Gnedin:1993,Bonanno:1994qh,Brady:1995ni,Brady:1994aq,Burko:1997fc,Burko:1997zy,Burko:1998jz,Husain:2000vm,Burko:2002qr,Burko:2002fv,MartinGarcia:2003gz,Dafermos:2004jp,Hansen:2005am,
Hod:1996ar,Hod:1998gy,Hod:1999wb,Sorkin:2000pc,Oren:2003gp,Dafermos:2003wr,Dafermos:2003yw},
which supports waves moving at the speed of light,
and is thus supposed to mimic gravitational radiation
while retaining the simplifying advantage of spherical symmetry.

Real astronomical black holes are however never isolated.
Supermassive black holes at the centers of galaxies
are thought to build up their observed
$\sim 10^6$--$10^{9} \unit{\Msun}$
masses
\cite{Kormendy:2001hb,Novak:2005km,Lauer:2006jg}
by gradual, albeit sporadic, accretion over
the age of the Universe.
Astronomical black holes,
whether supermassive or stellar-sized,
will continue to accrete
some mixture of baryons and dark matter from their galactic environments.
Even if nothing else is present,
a black hole will accrete cosmic microwave background photons.

Accretion provides an ongoing source of ingoing
and outgoing
matter inside a real black hole.
Matter free-falling from outside the horizon is necessarily initially ingoing,
but it may become outgoing inside the horizon
if it has enough charge (if the black hole is charged)
or enough angular momentum (if the black hole is rotating).
It seems likely that outgoing radiation
produced as a result of accretion
will soon overwhelm the rapidly decaying Price tail of outgoing radiation
produced by the initial collapse event.
This is especially true in a supermassive black hole,
whose mass acquired by accretion greatly exceeds
the mass of the original black hole formed by stellar collapse.

Occasional events of high accretion,
most notably for example from the merger of two black holes,
will re-energize a Price tail of outgoing radiation.
In a general situation of variable accretion, outgoing radiation will be
generated by some combination of ring-down from peak accretion events,
and steady accretion.

The purpose of this paper
is to clarify the physics of the mass inflation instability
inside spherical, charged black holes.
As originally proposed by \cite{Poisson:1990eh},
mass inflation is fueled by counter-streaming
ingoing and outgoing streams.
We take the point of view that
ingoing and outgoing streams are being generated continuously by accretion.
We treat the rates of generation of ingoing and outgoing streams
as free parameters,
and investigate how mass inflation depends on these rates.

In this paper we assume that ingoing and outgoing streams are non-interacting.
At typically low astronomical accretion rates,
the assumption of non-interacting streams is likely to be a good one,
at least in the realm of ``known physics.''
However, the assumption of non-interacting streams is likely to break
down at high accretion rates,
or at center-of-mass collision energies exceeding the Planck mass.
Interaction between ingoing and outgoing streams is likely to have
interesting consequences.
We hope to explore the situation of interacting streams
in a subsequent paper.

The strategy of this paper is to start from qualitative
and approximate arguments,
and to proceed to illustrative models
that support the qualitative conclusions.
To put the paper into context,
\S\ref{sgra}
presents the example of
the supermassive black hole at the center of our Galaxy.
To start the qualitative argument,
\S\ref{mechanism} identifies
the two Einstein equations~(\ref{dtbspherical})
that are at the heart of the mechanism of inflation.
Next, \S\ref{twoequalstreams}
presents the simplest possible example,
that of two symmetrically equal ingoing and outgoing streams,
each neutral, pressureless, and freely-falling.
The assumption of symmetrically equal ingoing and outgoing streams
is equivalent,
\S\ref{stationaryapproximation},
to the stationary approximation $\partial / \partial t = 0$,
previously introduced by
\cite{Burko:1997xa,Burko:1998az,Burko:1998jz},
who call it the homogeneous approximation
because the time direction $t$ is spacelike inside the horizon
(we follow the convention of \cite[p.~203]{Carroll:2004}
in referring to time translation symmetry as stationary,
even when the time direction is spacelike rather than timelike).
The assumption of equal streams is not realistic,
so \S\ref{twounequalstreams}
goes on to generalize to the case of two unequal streams.

The paper then moves on from approximations
to self-consistent, self-similar models of accreting,
charged, spherical black holes
\cite{Hamilton:2004av,Hamilton:2004aw}.
Section~\ref{baryondarkmatter}
considers two-stream baryon-plus-dark-matter models,
while \S\ref{singlefluid}
considers single-fluid models,
including the case of
a perfect fluid with an ultra-hard equation of state,
which can be considered as a model of a massless scalar field
\cite{Babichev:2008dy}.

All of this would be irrelevant if what happens during inflation
depends on events that occur in the far future
(as suggested by the Penrose diagram of the Reissner-Nordstr\"om geometry),
so \S\ref{future} discusses that issue.

Section~\ref{collisionrate}
checks whether the assumption of non-interacting streams is valid.

To conclude,
\S\ref{summary}
summarizes the findings of the paper,
and \S\ref{finalremarks}
appends some final remarks.


\section{An astronomical example}
\label{sgra}

It is useful to put the arguments of subsequent sections into context
by considering a particular example:
the $\sim 4 \times 10^6 \unit{\Msun}$
supermassive black hole
at the center of our Galaxy
\cite{Ghez:2003qj,
Eisenhauer:2005cv}.
In this section, results from subsequent sections
will simply be quoted without justification.

Mass inflation takes place at (or rather just above) the inner horizon.
Unless the black hole is rotating unusually slowly,
the inner horizon of the Milky Way black hole will be at a radius $r_-$ of
\begin{equation}
  r_-
  \sim
  G M / c^2
  \sim 10^7 \unit{km}
  \ ,
\end{equation}
or several times the radius of the Sun.

The characteristic dimensionless accretion rate $\mu$ of the black hole,
which can be thought of as the velocity with which the horizon is expanding
in units of the speed of light,
is expected to be of the order of the light crossing time of the black hole,
about 1 minute,
divided by the accretion timescale,
about the age of the Universe,
\begin{equation}
  \mu
  \sim
  {G M / c^3 \over \mbox{age of Universe}}
  \sim
  {1 \unit{min} \over 10^{10} \unit{yr}}
  \sim
  10^{-16}
  \ ,
\end{equation}
which is tiny.

Mass inflation is driven by relativistic counter-streaming
between ingoing and outgoing streams.
During inflation,
the counter-streaming velocity $u$ (meaning $\gamma v$)
exponentiates with an $\ee$-folding lengthscale of about $\mu r_-$
(eq.~(\ref{dlnbetadlnrmi}) or~(\ref{dlnbetadlnrmitwo})),
which is
\begin{equation}
\label{lsgra}
  \mu r_-
  \sim
  10^{-3} \unit{mm}
  \ .
\end{equation}
Thus as the counter-streaming ingoing and outgoing streams
drop 1 millimeter in radius,
the counter-streaming velocity increases by 1000 $\ee$-folds.
At this point, still at a radius of $\sim 10^{7} \unit{km}$,
the Weyl curvature,
center-of-mass energy density,
and such-like quantities easily exceed the Planck scale.

Presumably other physics of some kind intervenes,
whether it be quantum gravity,
or large collision cross-sections at super-Planck energies.
But if not, if ordinary general relativity continues to operate,
and if the streams remain non-interacting,
then the counter-streaming velocity would exponentiate up to
the absurdly huge value of
(eq.~(\ref{umaxcollapse}) or~(\ref{uamaxcollapse}))
\begin{equation}
\label{usgra}
  u
  \sim
  \ee^{1/\mu}
  \sim
  \ee^{10^{16}}
\end{equation}
before the inflationary period came to an end.

During inflation,
ingoing and outgoing streams both see each other highly blueshifted.
An outgoing observer sees ingoing material accreted from the external Universe
to the future of the time at which the observer jumped into the black hole,
while an ingoing observer sees outgoing material accreted from the external
Universe to the past of the time that they jumped in.
Despite the huge blueshift, each stream sees
only a modest amount of time go by on the other stream,
approximately one black hole crossing time per $\ee$-fold increase in blueshift
(eq.~(\ref{dtbmi})).
After 1000 $\ee$-folds of blueshift,
easily enough to pass the Planck scale,
each stream has seen less than one day to have passed on the other stream.
Thus what happens during inflation
depends only on the immediate future and past of the black hole,
not on its distant future or past.

From each of the ingoing and outgoing stream's own perspective,
all this happens in a blink,
approximately the light-crossing time across one $\ee$-folding length scale
of $10^{-3} \unit{mm}$,
which is a few femtoseconds ($10^{-15} \unit{s}$).
Hitting the inner horizon is like hitting the pavement
at the bottom of a skyscraper.
You sailed down, but then it's all over in an instant.

Although the center-of-mass energy density
of the counter-streaming ingoing and outgoing streams is vast,
the proper energy density of each stream individually remains modest,
comparable to its energy density before inflation.
Two cold streams being accelerated through
each other up to highly relativistic velocities
is characteristic of a particle accelerator.
The situation may aptly be termed
the ``black hole particle accelerator''
(\S\ref{collisionrate}).

\section{The mechanism of mass inflation}
\label{mechanism}

The purpose of this section
is to elucidate the physics of mass inflation sufficiently well
to provide a sound foundation for the approximations developed
in the following two sections,
\S\ref{twoequalstreams} and \S\ref{twounequalstreams}.
The arguments and approximations are intended to be simple,
but the reader should not be deceived into thinking
that it is trivial or obvious to derive approximations
that capture the behavior of inflation accurately.
To date,
the only simple approximation that has been considered in the literature
\cite{Burko:1997xa,Burko:1998az,Burko:1998jz,Hansen:2005am}
is the stationary approximation $\partial / \partial t = 0$,
which 
\cite{Burko:1997xa}
calls the homogeneous approximation
since the time direction is spacelike near the inner horizon.
As shown \S\ref{stationaryapproximation},
the stationary approximation is equivalent
to the (unrealistic) case of equal ingoing and outgoing streams.

This paper follows~\cite{Hamilton:2004av,Hamilton:2004aw}
in adopting the following line-element
for a general spherically symmetric spacetime,
in polar coordinates
$x^\mu \equiv \{ t , r , \theta , \phi \}$:
\begin{equation}
\label{metric}
  \dd s^2
  =
  - \,
  {\dd t^2 \over \alpha^2}
  +
  {1 \over \beta_r^2}
  \left(
  \dd r
  -
  \beta_t
  {\dd t \over \alpha}
  \right)^2
  +
  r^2
  ( \dd \theta^2 + \sin^2\!\theta \, \dd \phi^2 )
  \ .
\end{equation}
Here $r$ is the circumferential radius,
a gauge-invariant scalar,
defined such that the proper circumference
at radius $r$ is $2\pi r$.
As detailed in Appendix~\ref{sphericalspacetime},
the line-element~(\ref{metric})
encodes not only a metric,
but a complete orthonormal tetrad
$\bgamma_m \equiv \{ \bgamma_t , \bgamma_r , \bgamma_\theta , \bgamma_\phi \}$,
that is, a locally inertial frame, at each point of the spacetime.
The coefficients $\beta_t$ and $\beta_r$
in the line-element~(\ref{metric})
constitute the components of a tetrad-frame 4-vector,
the radial 4-gradient
\begin{equation}
\label{betam}
  \beta_m
  \equiv
  \partial_m r
  \ ,
\end{equation}
where
$\partial_m \equiv \bgamma_m \cdot \bpartial \equiv e_m{}^\mu \partial / \partial x^\mu$
denotes the directed derivative along the
$\bgamma_m$
tetrad axis
(not to be confused with the coordinate derivative
$\partial / \partial x^\mu$).
Physically,
the directed derivatives
$\partial_t$
and
$\partial_r$
are the proper time and radial derivatives
measured by a person at rest in the tetrad frame.
In particular,
$\beta_t \equiv \partial_t r$
equals the proper rate of change
of the circumferential radius $r$
measured by a person at rest in the tetrad frame
(in another notation,
this might be written $\dd r / \dd \tau$,
where $\tau$ is the proper time in the tetrad frame).
The scalar length squared of $\beta_m$
defines the interior,
or Misner-Sharp \cite{Misner:1964},
mass $M(r)$,
a gauge-invariant scalar,
by
\begin{equation}
\label{M}
  {2 M \over r}
  -
  1
  \equiv
  \beta^2
  \equiv
  - \,
  \beta_m \beta^m
  =
  \beta_t^2 - \beta_r^2
  \ .
\end{equation}
For reference,
Appendix~\ref{sphericalspacetime}
summarizes the various entities,
such as the tetrad connections and Einstein tensor,
derived from the line-element~(\ref{metric}).

Spherically symmetric spacetimes are described by
four independent Einstein equations.
Two of the four serve to enforce conservation of energy-momentum,
so if the equations governing the energy-momentum
of the system are arranged to satisfy conservation of energy-momentum,
as they should,
then any two of the Einstein equations
can be discarded as redundant.
For the two remaining Einstein equations,
it is most insightful to take the
expressions~(\ref{Grrspherical}) and (\ref{Gtrspherical})
for the components $G^{rr}$ and $G^{tr}$ of the Einstein tensor,
which yield the following Einstein equations for
the time evolution of $\beta_t$ and $\beta_r$:
\begin{subequations}
\label{dtbspherical}
\begin{align}
\label{dtbtspherical}
  \DD_t \beta_t
  &=
  - \,
  {M \over r^2}
  -
  4 \pi r p
  \ ,
\\
\label{dtbrspherical}
  \DD_t \beta_r
  &=
  4 \pi r f
  \ ,
\end{align}
\end{subequations}
where $\DD_m$ denotes the tetrad-frame covariant derivative,
$f \equiv T^{tr}$
is the energy flux,
and
$p \equiv T^{rr}$
is the radial pressure
measured in the tetrad frame.
Equations~(\ref{dtbspherical})
are valid in any (radially-moving) tetrad frame,
and for any form of energy-momentum;
the only requirement is that the spacetime be spherically symmetric.

The two Einstein equations~(\ref{dtbspherical})
are at the heart of mass inflation.
Equation~(\ref{dtbtspherical}) relates the radial acceleration
on the left hand side
to the gravitational force
on the right hand side.
The left hand side of equation~(\ref{dtbtspherical}),
$\DD_t \beta_t$,
is the proper acceleration of the circumferential radius $r$
measured by an observer who is in free-fall
and instantaneously at rest in the tetrad frame.
The right hand side of equation~(\ref{dtbtspherical})
is the gravitational force,
which consists of two terms,
the (apparently) Newtonian gravitational force $- M / r^2$,
and an additional term
$- 4 \pi r p$
proportional to the radial pressure $p$.

As will be seen below,
it is the second, pressure term,
in the gravitational force
on the right hand side of equation~(\ref{dtbtspherical})
that is the source of the fun,
first initiating mass inflation,
and then driving it exponentially.

Like the second half of a vaudeville act,
the second Einstein equation~(\ref{dtbrspherical})
also plays an indispensible role.
The quantity $\beta_r \equiv \partial_r r$
on the left hand side
is the proper radial gradient of the circumferential radius $r$
measured by a person at rest in the tetrad frame.
The sign of $\beta_r$
determines which way an observer at rest in the tetrad frame
thinks is ``outwards,''
the direction of larger circumferential radius $r$.
A positive $\beta_r$ means that
the observer thinks the outward direction points away from the black hole,
while a negative $\beta_r$ means that
the observer thinks the outward direction points towards from the black hole.
Outside the outer horizon
$\beta_r$ is necessarily positive,
because $\beta_m$ must be spacelike there.
But inside the horizon $\beta_r$ may be either positive or negative.
A tetrad frame can be defined as ``ingoing''
if the proper radial gradient $\beta_r$ is positive,
and ``outgoing'' if $\beta_r$ is negative.
In the Reissner-Nordstr\"om geometry,
ingoing geodesics have positive energy,
and outgoing geodesics have negative energy.
However,
the present definition of ingoing or outgoing based on the sign of $\beta_r$
is general -- there is no need for a timelike Killing vector
such as would be necessary to define the (conserved) energy of a geodesic.

Equation~(\ref{dtbrspherical})
shows that the proper rate of change
$\DD_t \beta_r$
in the radial gradient $\beta_r$
measured by an observer who is in free-fall
and instantaneously at rest in the tetrad frame
is proportional to the radial energy flux $f$
in that frame.
But ingoing observers tend to see energy flux pointing
away from the black hole,
while outgoing observers tend to see energy flux pointing
towards the black hole.
Thus the change in $\beta_r$
tends to be in the same direction as $\beta_r$,
amplifying $\beta_r$ whatever its sign.

\betadiagramfig

\subsection{Reissner-Nordstr\"om phase}

Figure~\ref{betadiagram}
illustrates how the two Einstein equations~(\ref{dtbspherical})
produce the three phases of mass inflation inside
a charged spherical black hole.

During the initial phase,
illustrated in the top panel of Figure~\ref{betadiagram},
the spacetime geometry is well-approximated
by the vacuum (more correctly, electrovac) Reissner-Nordstr\"om geometry.
During this phase
the radial energy flux $f$ is effectively zero,
so $\beta_r$ remains constant, according to equation~(\ref{dtbrspherical}).
The change in the infall velocity $\beta_t$,
equation~(\ref{dtbtspherical}),
depends on the competition between the Newtonian gravitational force
$- M / r^2$,
which is always attractive
(tending to make the infall velocity $\beta_t$ more negative),
and the gravitational force $- 4 \pi r p$ sourced by the radial pressure $p$.
In the Reissner-Nordstr\"om geometry,
the static electric field produces a negative radial pressure, or tension,
$p = - Q^2 / ( 8 \pi r^4 )$,
which produces a gravitational repulsion
$- 4 \pi r p = Q^2 / ( 2 r^3 )$.
At some point
(depending on the charge-to-mass ratio $Q/M$)
inside the outer horizon,
the gravitational repulsion produced by the tension of the electric field
exceeds the attraction produced by the interior mass $M$,
so that the infall velocity $\beta_t$ slows down.
This regime,
where the (negative) infall velocity $\beta_t$
is slowing down (becoming less negative),
while $\beta_r$ remains constant,
is illustrated in the top panel of Figure~\ref{betadiagram}.

If the initial Reissner-Nordstr\"om phase were to continue,
then the radial 4-gradient $\beta_m$ would become lightlike.
In the Reissner-Nordstr\"om geometry
this does in fact happen,
and where it happens defines the inner horizon.
The problem with this is that
the lightlike 4-vector $\beta_m$ points in one direction for ingoing frames,
and in the opposite direction for outgoing frames.
If $\beta_m$ becomes lightlike,
then ingoing and outgoing frames are streaming through each other
at the speed of light.
This is the infinite blueshift at the inner horizon first pointed out by
\cite{Penrose:1968}.

If there were no matter present,
or if there were only one stream of matter,
either ingoing or outgoing but not both,
then $\beta_m$ could indeed become lightlike.
But if both ingoing and outgoing matter are present,
even in the tiniest amount,
then it is physically impossible for the ingoing and outgoing frames
to stream through each other at the speed of light.

If both ingoing and outgoing streams are present,
then as they race through each other ever faster,
they generate a radial pressure $p$,
and an energy flux $f$,
which begin to take over as the main source
on the right hand side of the Einstein equations~(\ref{dtbspherical}).
This is how mass inflation is ignited.

\subsection{Inflationary phase}

The infalling matter now enters the second,
mass inflationary phase,
illustrated in the middle panel of Figure~\ref{betadiagram}.

During this phase,
the gravitational force on the right hand side of
the Einstein equation~(\ref{dtbtspherical})
is dominated by the pressure $p$
produced by the counter-streaming ingoing and outgoing matter.
The mass $M$ is completely sub-dominant during this phase
(in this respect, the designation ``mass inflation'' is misleading,
since although the mass inflates, it does not drive inflation).
The counter-streaming pressure $p$ is positive,
and so accelerates the infall velocity $\beta_t$
(makes it more negative).
At the same time,
the radial gradient $\beta_r$
is being driven by the energy flux $f$,
equation~(\ref{dtbrspherical}).
For typically low accretion rates,
the streams are cold,
in the sense that the streaming energy density
greatly exceeds the thermal energy density,
even if the accreted material is at relativistic temperatures.
This follows from the fact that
for mass inflation to begin,
the gravitational force produced by
the counter-streaming pressure $p$
must become comparable to that produced by the mass $M$,
which for streams of low proper density
requires a hyper-relativistic streaming velocity.
For a cold stream of proper density $\rho$
moving at 4-velocity $u^m \equiv \{ u^t , u^r , 0 , 0 \}$,
the streaming energy flux would be $f \sim \rho u^t u^r$,
while the streaming pressure would be $p \sim \rho (u^r)^2$.
Thus their ratio
$f / p \sim u^t / u^r$
is slightly greater than one.
It follows that,
as illustrated in the middle panel of Figure~\ref{betadiagram},
the change in $\beta_r$ slightly exceeds the change in $\beta_t$,
which drives the 4-vector $\beta_m$,
already nearly lightlike,
to be even more nearly lightlike.
This is mass inflation.

Inflation feeds on itself.
The radial pressure $p$ and energy flux $f$ generated by
the counter-streaming ingoing and outgoing streams
increase the gravitational force.
But, as illustrated in the middle panel of Figure~\ref{betadiagram},
the gravitational force acts in opposite directions
for ingoing and outgoing streams,
tending to accelerate the streams faster through each other.
As
\cite{Hamilton:2004aw}
put it,
the gravitational force is always inwards,
meaning in the direction of smaller radius,
but the inward direction is towards the black hole for ingoing streams,
and away from the black hole for outgoing streams.

The feedback loop
in which the streaming pressure and flux
increase the gravitational force,
which accelerates the streams faster through each other,
which increases the streaming pressure and flux,
is what drives mass inflation.
Inflation produces an exponential growth
in the streaming energy,
and along with it the interior mass, and the Weyl curvature.

\subsection{Collapse phase}

It might seem that inflation
is locked into an exponential growth from which there is no exit.
But the Einstein equations~(\ref{dtbspherical})
have one more trick up their sleave.

For the counter-streaming velocity to continue to increase
requires that the change in $\beta_r$ from equation~(\ref{dtbrspherical})
continues to exceed the change in $\beta_t$ from equation~(\ref{dtbtspherical}).
This remains true
as long as the counter-streaming pressure $p$ and energy flux $f$
continue to dominate the source
on the right hand side of the equations.
But the mass term $- M / r^2$ also makes a contribution
to the change in $\beta_t$,
equation~(\ref{dtbtspherical}).
As will be seen in the examples of the next two sections,
\S\S\ref{twoequalstreams} and \ref{twounequalstreams},
at least in the case of pressureless streams
the mass term exponentiates slightly faster than the pressure term.
At a certain point,
the additional acceleration produced by the mass
means that the combined gravitational force
$M/r^2 + 4 \pi r p$
exceeds $4 \pi r f$.
Once this happens,
the 4-vector $\beta_m$,
instead of being driven to becoming more lightlike,
starts to become less lightlike.
That is, the counter-streaming velocity starts to slow.
At that point inflation ceases,
and the streams quickly collapse to zero radius.

It is ironic that it is the increase of mass that brings mass inflation
to an end.
Not only does mass not drive mass inflation,
but as soon as mass begins to contribute significantly to the
gravitational force, it brings mass inflation to an end.

\section{Two equal streams}
\label{twoequalstreams}

The purpose of this section is to present
the simplest possible quantitative example
of how the counter-streaming instability described
qualitatively in the previous section works.
The model is that of two equal ingoing and outgoing streams
of neutral, pressureless matter.
As shown in
\S\ref{stationaryapproximation},
the approximation is equivalent to the stationary
(homogeneous) approximation first considered by
\cite{Burko:1997xa,Burko:1998az,Burko:1998jz}.
The assumed symmetry between ingoing and outgoing streams is not realistic,
but has the virtue of simplicity.
A more realistic but still idealized model with two unequal streams
is solved in the next section, \S\ref{twounequalstreams}.
Self-consistent models
are presented in later sections.

For pedagogy,
the details of the solution for two equal streams
are left as a problem for the reader,
Appendix~\ref{problem}.
This problem appeared on the final exam of the first author's
graduate course on general relativity in Spring 2008.
The problem includes sufficient hints that the reader should
be able to solve it.
In any case,
elements of the solution are given below.

As seen in the previous section, \S\ref{mechanism},
the key object is the radial 4-gradient $\beta_m$,
equation~(\ref{betam}).
The radial 4-gradient contains two degrees of freedom,
its time and radial components $\beta_t$ and $\beta_r$.
It is convenient to recast these two degrees of freedom in terms
of a magnitude $\beta \equiv \lvert \beta_m \rvert$
and the radial component $u \equiv u^r$
of a radial 4-velocity $u^m$
\begin{equation}
\label{betamu}
  \beta_m
  \equiv
  \{ \beta_t , \beta_r , 0 , 0 \}
  \equiv
  -
  \beta
  \{ \sqrt{1 + u^2} , u , 0 , 0 \}
  \ .
\end{equation}
The magnitude $\beta$, which is positive,
is related to the interior mass $M$ by
$\beta^2 = 2 M / r - 1$,
equation~(\ref{M}).
The 4-velocity $u^m \equiv \{ \sqrt{1 + u^2} , u , 0 , 0 \}$
is the 4-velocity
of the stream relative to the no-going frame
at the boundary between ingoing and outgoing,
where $\beta_r = 0$.
The sign of $u \equiv u^r$ is negative for ingoing, positive for outgoing.

The central equation that emerges from the analysis in Appendix~\ref{problem} is
[the following equation repeats eq.~(\ref{dlnbdlnusym})]
\begin{equation}
\label{dlnbdlnusymtext}
  {\dd \ln \beta \over \dd \ln u}
  =
  {
  - \, \lambda + \beta^2 + \mu u^2
  \over
  \lambda - \beta^2 + \mu + \mu u^2
  }
  \ .
\end{equation}
The dimensionless parameters
$\lambda$
and
$\mu$
in equation~(\ref{dlnbdlnusymtext})
are well-approximated as constant during mass inflation.
The dimensionless parameter $\lambda$ is
\begin{equation}
\label{lambda}
  \lambda
  \equiv
  Q^2 / r^2 - 1
  \ ,
\end{equation}
where $Q$ is the interior charge of the black hole
and $r$ is the radius.
It is a feature of inflation that during it the radius $r$ hardly changes,
a fact confirmed below, equation~(\ref{dlnbetadlnrmi}).
The radius may be approximated by
the radius of the inner horizon
$r \approx r_- = M - \sqrt{ M^2 - Q^2 }$
of the parent Reissner-Nordstr\"om black hole.
The parameter $\lambda$ is typically a number of order unity;
for example,
if $Q / M = 0.8$, then $\lambda = 3$.
The dimensionless parameter $\mu$
in equation~(\ref{dlnbdlnusymtext}) is
\begin{equation}
\label{mu}
  \mu
  \equiv
  16 \pi r^2 \rho
  \ ,
\end{equation}
proportional to the product of the radius $r$ squared
and the proper mass-energy density $\rho$
of either of the ingoing or outgoing streams.
Perhaps surprisingly,
the proper density $\rho$ of either stream
in its own frame remains almost constant during inflation.
What exponentiates during inflation is the counter-streaming velocity,
and along with it the center-of-mass energy density;
by comparison the energy density of each stream changes little,
because its volume changes little.
The fact that a volume element remains little distorted
during inflation even though the tidal force,
as measured by the Weyl curvature scalar,
exponentiates to huge values
was first pointed out by \cite{Ori:1991}.
The physical reason for the small tidal distortion
despite the huge tidal force
is that the proper time over which the force operates is tiny.
The parameter $\mu$
is proportional to the accretion rate of the black hole,
and for simplicity
$\mu$ is referred to loosely in this paper as the ``accretion rate.''
As seen in \S\ref{sgra},
astronomical accretion rates are typically small (tiny),
so $\mu$ is small (tiny).

Equation~(\ref{dlnbdlnusymtext})
is a one-dimensional ordinary differential equation,
whose solution defines curves in the $\beta$--$u$ plane.
The behavior depends on the relative magnitudes of three terms,
$\lambda$,
$\mu u^2$,
and $\beta^2$
(a fourth term $\mu$ is never important)
that appear in the numerator and denominator
on the right hand side of equation~(\ref{dlnbdlnusymtext}).
The three phases of evolution
described in \S\ref{mechanism}
occur as each of the three terms dominates,
Table~\ref{phase}.
\phasetab
As listed in the middle column of Table~\ref{phase},
physically,
the dominance of each term
expresses the dominance of three sources
of the gravitational force on the right hand side
of the Einstein equations~(\ref{dtbspherical}):
$\lambda$ dominates
if the negative pressure of the electric field dominates;
$\mu u^2$
dominates if
the counter-streaming pressure and flux dominate;
and
$\beta^2$
dominates if the interior mass $M$ dominates.
The three phases are considered in the next three subsections.

\subsection{Reissner-Nordstr\"om phase}
\label{rnsym}

Because the accretion rate $\mu$ is small,
initially the streaming $\mu u^2$ term (and the $\mu$ term)
in equation~(\ref{dlnbdlnusymtext})
can be ignored.
During this initial, Reissner-Nordstr\"om, phase,
equation~(\ref{dlnbdlnusymtext}) simplifies to
\begin{equation}
\label{dlnbdlnurntext}
  {\dd \ln \beta \over \dd \ln u}
  =
  {
  - \, \lambda + \beta^2
  \over
  \lambda - \beta^2
  }
  =
  -1
  \ .
\end{equation}
The initial values of $\beta$ and $u$
are set by the boundary conditions of accretion,
and are generically of order unity;
their precise initial values are not important to the argument.
The solution of equation~(\ref{dlnbdlnurntext}) is
\begin{equation}
\label{betarntext}
  \beta
  =
  {C \over u}
\end{equation}
where $C$ is a constant of order unity,
set by the boundary conditions.
This Reissner-Nordstr\"om phase of evolution is labeled ``RN''
in the $\beta$-$u$ diagram of Figure~\ref{budiagram}.

\budiagramfig

In the Reissner-Nordstr\"om geometry,
$\beta$ would decrease to zero at the inner horizon,
while the velocity $u$ would diverge,
in accordance with equation~(\ref{betarntext}).
This reflects the tendency of the 4-vector $\beta_m$
to become lightlike at the inner horizon,
as described in \S\ref{mechanism}
and illustrated in the top panel of Figure~\ref{betadiagram}.

As the velocity $u$ becomes larger on the approach to the inner horizon,
the term $\mu u^2$ in equation~(\ref{dlnbdlnusymtext}),
which is proportional to the streaming pressure
in the no-going (= center-of-mass, here) frame,
eventually starts to become appreciable,
however tiny the accretion rate $\mu$ may be initially.
The Reissner-Nordstr\"om phase ends when
$\mu u^2 \approx \lambda$,
which corresponds to the streaming pressure and flux
taking over as the dominant source of gravity on the right hand side
of the Einstein equations~(\ref{dtbspherical}).

\subsection{Inflationary phase}
\label{inflationsym}

During the Reissner-Nordstr\"om phase,
$\beta$ is driven to a small value,
while the velocity $u$ increases to the point
where the streaming pressure begins to take over.
Since the $\beta^2$ terms in equation~(\ref{dlnbdlnusymtext})
are now small, the equation may be approximated by
\begin{equation}
\label{dlnbdlnumitext}
  {\dd \ln \beta \over \dd \ln u}
  =
  {
  - \, \lambda + \mu u^2
  \over
  \lambda + \mu u^2
  }
  \ .
\end{equation}
The solution of equation~(\ref{dlnbdlnumitext})
that initially matches the Reissner-Nordstr\"om solution~(\ref{betarntext})
is
\begin{equation}
\label{betamitext}
  \beta
  =
  {C ( \lambda + \mu u^2 ) \over \lambda u}
  \ .
\end{equation}
Equation~(\ref{dlnbdlnumitext})
shows that $\beta$
reaches a minimum at $u^2 = \lambda / \mu$,
where, according to the solution~(\ref{betamitext}),
\begin{equation}
\label{betaminmi}
  \beta
  =
  2 \lvert C \rvert \sqrt{ \mu / \lambda }
  \ .
\end{equation}
The solution then enters the inflationary phase,
where
$\beta$ increases in proportion to $u$
\begin{equation}
\label{betamilate}
  \beta
  =
  {C \mu u \over \lambda}
  \quad
  \mbox{for $u^2 \gg \lambda / \mu$}
  \ .
\end{equation}
This phase is labeled ``inflation''
in the $\beta$-$u$ diagram of Figure~\ref{budiagram}.

That something dramatic is happening
can be seen from the equations~(\ref{dlnbudlnrsym})
relating the change in $\beta$ and $u$
to the change in the radius $r$.
During the inflationary phase where the $\mu u^2$ term dominates,
the inflationary growth rate from equations~(\ref{dlnbudlnrsym}) is
\begin{equation}
\label{dlnbetadlnrmi}
  {\dd \ln \beta \over \dd \ln r}
  =
  {\dd \ln u \over \dd \ln r}
  =
  - \, {\lambda^2 \over 2 C^2 \mu}
  \ .
\end{equation}
Since $\mu$ is small (tiny),
these rates are large (huge).
That is,
$\beta$ and $u$ increase dramatically
while the radius $r$ decreases only a small amount.

Equation~(\ref{dlnbetadlnrmi})
shows that the smaller the accretion rate $\mu$,
the larger the inflationary growth rate.
This counter-intuitive feature is confirmed in the self-consistent models
described in \S\S\ref{baryondarkmatter} and \ref{singlefluid}.

\subsection{Collapse phase}
\label{collapsesym}

During the inflationary phase
the streaming term $\mu u^2$ in equation~(\ref{dlnbdlnusymtext})
is driven to a huge value, dwarfing the $\lambda$ term.
But $\beta$ has also been increasing during inflation,
and the $\beta^2$ mass term in equation~(\ref{dlnbdlnusymtext})
begins to make its presence felt.
With the $\mu u^2$ and $\beta^2$ terms included
but the $\lambda$ terms dropped,
equation~(\ref{dlnbdlnusymtext}) is approximately
\begin{equation}
\label{dlnbdlnucollapse}
  {\dd \ln \beta \over \dd \ln u}
  =
  {
  \beta^2 + \mu u^2
  \over
  - \beta^2 + \mu u^2
  }
  \ .
\end{equation}
The right hand side of equation~(\ref{dlnbdlnucollapse})
is slightly greater than 1,
which shows that $\beta$ grows slightly faster than $u$.
Eventually,
the $\beta^2$ term grows to the point that it equals $\mu u^2$.
At this point the denominator of the right hand side of
equation~(\ref{dlnbdlnucollapse}) is zero.
The velocity $u$ then begins to decrease,
while $\beta$ continues to increase.
This signals the end of mass inflation.

Equation~(\ref{dlnbdlnucollapse}) has an exact solution given by
the implicit equation
\begin{equation}
\label{ubcollapse}
  \beta u
  =
  {\lambda^2 \over C \mu} 
  \exp \left(
  {\lambda^2 \over 2 C^2 \mu}
  \right)
  \exp \left(
  -
  {\mu u^2 \over 2 \beta^2}
  \right)
  \ ,
\end{equation}
in which the constant integration factor
$\lambda^2 / ( C \mu ) \ee^{\lambda^2 / ( 2 C^2 \mu )}$
comes from matching to the inflationary solution~(\ref{betamilate})
at $\beta^2 = \lambda$,
where
equation~(\ref{dlnbdlnucollapse})
takes over from
equation~(\ref{dlnbdlnumitext}).
Equation~(\ref{dlnbdlnucollapse})
shows that the velocity $u$ reaches a maximum
when $\beta^2 = \mu u^2$,
at which point equation~(\ref{ubcollapse}) shows that
\begin{equation}
\label{umaxcollapse}
  u
  =
  {\lambda \over C^{1/2} \mu^{3/4}}
  \exp \left(
  {\lambda^2 \over 4 C^2 \mu}
  -
  {1 \over 4}
  \right)
  \ .
\end{equation}
Since the accretion rate $\mu$ is small,
the maximum value of the velocity $u$ is exponentially vast,
the exponential of a large number,
$u \sim \ee^{1 / \mu}$.
In realistic situations,
the curvature and center-of-mass energy
are likely to exceed the Planck scale before
this vast value is attained,
and quantum gravity will presumably intervene.

However,
if the accretion rate $\mu$ is large enough,
as may happen during the first blush of gravitational collapse
of a black hole,
or during a black hole merger,
then the maximum value~(\ref{umaxcollapse})
may be attained before quantum gravity becomes important.
It was argued in \S\ref{inflationsym}
that the radius $r$ decreases only slightly during mass inflation.
However,
the number of $\ee$-folds that elapse in attaining
the peak velocity~(\ref{umaxcollapse}) is large enough that
the radius drops appreciably.
Specifically,
in reaching the maximum velocity~(\ref{umaxcollapse})
the radius $r$ decreases by approximately
\begin{equation}
  \Delta \ln r
  \approx
  -
  {2 C^2 \mu \over \lambda^2} \Delta \ln u
  \approx
  - \frac{1}{2}
  \ ,
\end{equation}
or a factor of $\ee^{-1/2} \approx 0.6$.

After the velocity $u$ peaks out,
the matter collapses precipitously to zero radius.
After the peak, equation~(\ref{ubcollapse}) simplifies to
\begin{equation}
  \beta u
  =
  {\lambda^2 \over C \mu} 
  \exp \left(
  {\lambda^2 \over 2 C^2 \mu}
  \right)
  \ ,
\end{equation}
which is just a constant.
This phase is labeled ``collapse''
in the $\beta$-$u$ diagram of Figure~\ref{budiagram}.
During the collapse phase,
the $\beta^2$ mass term dominates,
and it follows from equation~(\ref{dlnbdlnrsym})
that $\beta$ varies with radius $r$ as
\begin{equation}
\label{dlnbdlnrcollapse}
  {\dd \ln \beta \over \dd \ln r}
  \approx
  - \,
  {1 \over 2}
  \ .
\end{equation}
This shows that the radius $r$ is no longer stuck
at its inflationary value.
The proper radial infall velocity
$\beta_t \approx - \beta u$
is exponentially vast,
and the streams of matter plunge to zero radius in an instant.
Since the interior mass $M$ is related to $\beta$
by
$2 M / r - 1 = \beta^2$,
and $\beta$ is exponentially vast,
equation~(\ref{dlnbdlnrcollapse}) shows that
\begin{equation}
  {\dd \ln M \over \dd \ln r}
  =
  0
  \ .
\end{equation}
In other words, the interior mass $M$ stops increasing.
The value of $M$ attained
as the radius drops to zero
is the exponentially huge value
\begin{equation}
\label{Mcollapse}
  M
  \approx
  {\lambda r_- \over 2}
  \exp \left(
  {\lambda^2 \over 2 C^2 \mu}
  \right)
  \ ,
\end{equation}
where $r_-$,
the radius of the inner horizon of the original parent
Reissner-Nordstr\"om black hole,
is the approximate value of the radius during inflation.

\subsection{Stationary approximation}
\label{stationaryapproximation}

It might seem that
in the limit of small accretion rates
it would be reasonable to assume the stationary approximation
\begin{equation}
\label{stationary}
  {\partial \over \partial t}
  \equiv 0
  \ ,
\end{equation}
since as seen from outside the black hole hardly changes as time goes by.
However,
as will now be shown,
the stationary approximation~(\ref{stationary})
can be applied consistently to mass inflation
only in the (unrealistic) case
of symmetrically equal ingoing and outgoing streams,
which is the case considered in this section.
The approximation is termed
the homogeneous approximation by
\cite{Burko:1997xa,Burko:1998az,Burko:1998jz}
because
the Killing vector associated with time translation symmetry
is spacelike inside the black hole.
We follow \cite[p.~203]{Carroll:2004}
in referring to the approximation as stationary,
since it is just a single global symmetry of the spacetime.
The conclusions of this section are consistent with those of
\cite{Burko:1997xa,Burko:1998az,Burko:1998jz,Hansen:2005am},
who emphasize that the stationary (homogeneous) approximation
has the virtue of simplicity,
and captures the general behavior of mass inflation,
but does not predict accurately the behavior of more realistic models,
even in the limit of vanishingly small accretion rates.

The implications of the stationary approximation~(\ref{stationary})
can be drawn from the following general identity,
valid in arbitrary spherically symmetric spacetimes:
\begin{equation}
\label{Mdot}
  \beta_t \beta_r
  \left(
  G_{tt}
  +
  G_{rr}
  \right)
  -
  \left(
  \beta_r^2 + \beta_t^2
  \right)
  G_{tr}
  =
  -
  {2 \alpha \beta_r \over r^2}
  {\partial M \over \partial t}
  \ .
\end{equation}
In the stationary approximation~(\ref{stationary}),
the right hand side of equation~(\ref{Mdot}) vanishes,
and equation~(\ref{Mdot})
then implies that in the center-of-mass frame, where $G_{tr} = 0$,
one of the following three possibilities applies:
\begin{equation}
\label{zeroMdotcom}
  \begin{array}{rl}
  \mbox{either~~} & \beta_t = 0
  \ ,
  \\
  \mbox{or~~} & \beta_r = 0
  \ ,
  \\
  \mbox{or~~} & \rho + p = 0
  \ ,
  \end{array}
\end{equation}
where $\rho$ and $p$
denote the proper center-of-mass density and radial pressure.
Two of the three possibilities~(\ref{zeroMdotcom}),
namely $\beta_t = 0$ or $\rho + p = 0$,
are incompatible with mass inflation.
In the superluminally infalling inflationary zone,
$\beta_m$ is timelike,
$\beta_t^2 > \beta_r^2$,
and $\beta_t$ is strictly negative;
indeed,
$\beta_t^2 - \beta_r^2 = 2 M / r - 1$
exponentiates to huge values during inflation,
so $\beta_t$ is hugely negative.
Similarly,
the proper center-of-mass density and radial pressure
exponentiate to huge values during inflation,
so $\rho + p$ is hugely positive, not zero.

This leaves the third possibility,
that $\beta_r = 0$ in the center-of-mass frame
(note that $\alpha \beta_r$ remains finite when $\beta_r = 0$).
The possibility corresponds to the case of symmetrically
equal ingoing and outgoing streams,
where by symmetry the center-of-mass frame
is the no-going frame, $\beta_r = 0$,
at the boundary between ingoing and outgoing.
As
\cite{Hansen:2005am}
put it,
the stationary (homogeneous)
approximation applied to the case of a massless scalar
field implies that
``the scalar field can be represented as a sum
of two equal fluxes moving in opposite directions
along the (spacelike) $t$ axis.''
\cite{Hansen:2005am}
also apply the homogeneous approximation
to the case of single perfect fluids
with either pressureless ($p = 0$) or relativistic ($p/\rho = 1/3$)
equations of state.
However,
these single-fluid cases are irredeemably unrealistic,
since they suppress the inflation that would occur
if even the tiniest admixtures of ingoing and outgoing parts
were admitted.

The stationary approximation~(\ref{stationary})
correctly reproduces the qualitative behavior of mass inflation,
but does not capture accurately what happens
in the more realistic case of unequal streams,
\S\ref{twounequalstreams}.
If the stationary approximation~(\ref{stationary})
were true,
then geodesics would be characterized by
a conserved energy (per unit mass) $E$,
given by the covariant time component $\upsilon_t$
of the coordinate-frame 4-velocity $\upsilon^\mu$
(the coordinate-frame 4-velocity $\upsilon^\mu$
is written with a Greek upsilon
to avoid any possible confusion with
the tetrad-frame 4-velocity $u^m$,
written with a Latin u).
The ratio $E_b / E_a$ of conserved energies
of freely-falling streams $b$ and $a$,
expressed in the tetrad-frame of freely-falling stream $a$, is
\begin{equation}
\label{stationaryE}
  {E_b \over E_a}
  =
  {\upsilon_{b,t} \over \upsilon_{a,t}}
  =
  {e^m{}_{t} \, \eta_{mn} u_{ba}^n \over e^t{}_t}
  =
  u_{ba}^t + {\beta_{a,t} \over \beta_{a,r}} u_{ba}^r
  =
  {u_b \over u_a}
  \ .
\end{equation}
In the situation considered in \S\ref{twounequalstreams}
of two unequal streams accreting at constant rates,
the ratio $E_b/E_a$ of energies should be constant,
and then equation~(\ref{stationaryE})
would imply that
$u_b / u_a$
is constant.
But in fact the inflationary solution shows that
$u_b / u_a$ is not constant.
Rather it varies from
$u_b / u_a = C_b / C_a$
in the Reissner-Nordstr\"om phase,
equation~(\ref{betaa}),
to
$u_b / u_a = C_a \mu_a / ( C_b \mu_b )$
in the inflationary phase,
equation~(\ref{betaamilate}),
to
\begin{equation}
  {u_b \over u_a}
  \approx
  {C_a \mu_a \over C_b \mu_b}
  \exp \left[ {\lambda^2 ( C_b^2 \mu_b - C_a^2 \mu_a ) \over 2 C_a^2 \mu_a C_b^2 \mu_b} \right]
\end{equation}
in the collapse phase,
equation~(\ref{ubacollapse})
in the limit of small accretion rates
and late times,
$u_a \ll \beta$.

\section{Two unequal streams}
\label{twounequalstreams}

The situation of symmetrically equal ingoing and outgoing streams
considered in the previous section, \S\ref{twoequalstreams},
is straightforward to solve.
However,
the situation is not realistic,
since in reality the mix of ingoing and outgoing streams
depends on boundary conditions
outside the outer horizon
that are unlikely to produce equal streams.

This section considers the more realistic situation of two unequal streams.
The case remains analytically solvable provided that,
as in \S\ref{twoequalstreams},
the streams are idealized as neutral and pressureless,
and certain slowly-varying parameters
($\lambda$ and $\mu$)
are again approximated as constants for each stream.
The analytic solution for unequal streams can be compared to
results from self-consistent models of accreting black holes,
and it will be seen that the analytic solution
captures accurately what happens
in the self-consistent models.


For two unequal streams,
equation~(\ref{dlnbdlnusymtext}) generalizes to the two equations
\begin{equation}
\label{dlnbdlnutwo}
  {\dd \ln \beta \over \dd \ln u_a}
  =
  {
  - \, \lambda + \beta^2 + \mu_b u_b^2
  \over
  \lambda - \beta^2 + \mu_b u_b^2
  }
  \ ,
\end{equation}
in which $a$ represents either of the ingoing or outgoing streams,
and $b$ represents the other stream.
Equation~(\ref{dlnbdlnutwo}) is the ratio of
equations~(\ref{dlnbdlnrtwo}) and (\ref{dlnbdlnrtwo})
derived in Appendix~\ref{multiple}.
The accretion rate $\mu_b$ of stream $b$
in equation~(\ref{dlnbdlnutwo}) is given by
equation~(\ref{mu}), but subscripted,
\begin{equation}
\label{mub}
  \mu_b
  \equiv
  16 \pi r^2 \rho_b
  \ .
\end{equation}
Being a gauge-invariant scalar,
$\beta$ in equation~(\ref{dlnbdlnutwo})
is the same for both streams.

As in the case of equal streams considered in \S\ref{twoequalstreams},
the infalling matter goes through three phases of evolution,
Table~\ref{phase},
which are considered in turn in the next three subsections.

\subsection{Reissner-Nordstr\"om phase}

Realistically, the accretion rates $\mu_a$ of both
ingoing and outgoing streams are likely to be small
(even if one is much larger than the other),
so the $\mu_b u_b^2$ terms
on the right hand side of equation~(\ref{dlnbdlnutwo}) can be ignored.
This is the initial Reissner-Nordstr\"om phase,
and during it equation~(\ref{dlnbdlnutwo}) for each stream
is well-approximated by
\begin{equation}
\label{dlnbdlnuarn}
  {\dd \ln \beta \over \dd \ln u_a}
  =
  {
  - \, \lambda + \beta^2
  \over
  \lambda - \beta^2
  }
  =
  -1
  \ .
\end{equation}
Equation~(\ref{dlnbdlnuarn}) solves to
\begin{equation}
\label{betaa}
  \beta
  =
  {C_a \over u_a}
\end{equation}
for each stream,
where $C_a$ is a constant,
which will generally be different for each stream.
Note that
$C_a$ has the same sign as $u_a$,
which is negative ingoing, positive outgoing.

\subsection{Inflationary phase}
\label{inflationtwo}

As in the case of equal streams,
$\beta$ is driven to a small value during the Reissner-Nordstr\"om phase,
while the velocities $u_a$ of both streams increase,
equation~(\ref{betaa}).
Eventually the $\mu_b u_b^2$ streaming terms
on the right hand side of equation~(\ref{dlnbdlnutwo})
become important,
even though the accretion rate $\mu_b$ may be tiny.
Since the $\beta^2$ mass terms in equation~(\ref{dlnbdlnutwo}) are
by that time small,
equation~(\ref{dlnbdlnutwo}) becomes
\begin{equation}
\label{dlnbdlnuami}
  {\dd \ln \beta \over \dd \ln u_a}
  =
  {
  - \, \lambda + \mu_b u_b^2
  \over
  \lambda + \mu_b u_b^2
  }
\end{equation}
where as before $a$ represents either of the two streams,
while $b$ represents the other.
Equation~(\ref{dlnbdlnuami})
is a pair of coupled ordinary differential equations.
The solution
matching the Reissner-Nordstr\"om initial conditions~(\ref{betaa}) is
\begin{equation}
\label{betaami}
  \beta
  =
  {\lvert C_a \rvert ( \lambda + \mu_a u_a^2 )
  + \sqrt{ \bigl[ C_a ( \lambda - \mu_a u_a^2 ) \bigr]^2
  + \lambda \mu_b ( 2 C_b u_a )^2 }
  \over 2 \lambda \lvert u_a \rvert}
\end{equation}
for either stream.
Equation~(\ref{dlnbdlnuami})
shows that $\beta$ goes through a minimum
when $\mu_b u_b^2 = \lambda$,
at which point $\beta$ is,
according to the solution~(\ref{betaami}),
\begin{equation}
  \beta
  =
  {\lvert C_a \rvert \sqrt{\mu_a} + \lvert C_b \rvert \sqrt{\mu_b}
  \over \sqrt{\lambda}}
  \ ,
\end{equation}
which may be compared to the equal stream result~(\ref{betaminmi}).
Given the solution~(\ref{betaami}),
integration of equations~(\ref{dlnbudlnrtwo})
yields the radius $r_a$ of each stream:
\begin{equation}
\label{rami}
  r_a
  \propto
  \lvert u_a u_b \rvert^{- C_b^2 \mu_b / \lambda^2}
  \,
  \ee^{C_b \beta / ( \lambda u_b )}
  \ ,
\end{equation}
where the constant of proportionality may be set, for example,
by the value of $r_a$ where $\beta$ goes through its minimum.
Note that the radii $r_a$ are (slightly)
different for each stream,
since each follows a different trajectory into the black hole.
If $u_a$ and $u_b$ are expressed in terms of $\beta$
through equations~(\ref{betaami}),
then equation~(\ref{rami}) gives the radius $r_a$ of stream $a$
as a function of $\beta$ along its worldline.

The solution then enters the inflationary phase,
where $\beta$ and the velocities $u_a$ and $u_b$ increase
in proportion to each other
\begin{equation}
\label{betaamilate}
  \beta
  =
  {C_a \mu_a u_a \over \lambda}
  \ .
\end{equation}
The inflationary growth rate of each stream is,
from equations~(\ref{dlnbudlnrtwo}),
\begin{equation}
\label{dlnbetadlnrmitwo}
  {\dd \ln \beta \over \dd \ln r_a}
  =
  {\dd \ln u_a \over \dd \ln r_a}
  =
  - \, {\lambda^2 \over 2 C_b^2 \mu_b}
  \ ,
\end{equation}
where as above $a$ represents either of the two streams,
and $b$ the other.

\bufig

Figure~\ref{bu}
compares the analytic solution~(\ref{betaami})
to a self-consistent self-similar solution
of the kind described by \cite{Hamilton:2004aw}
and explored further in \S\ref{baryondarkmatter}.
For the model shown,
the boundary conditions at the outer sonic point are:
black hole charge-to-mass
$\Qbh/\Mbh = 0.8$,
mass accretion rate
$\Mbhdot = 0.01$,
baryonic equation of state
$w \equiv p_b / \rho_b = 10^{-6}$,
and proper dark-matter-to-baryon density
$\rho_d / \rho_b = 0.001$,
as listed in Table~\ref{par}.
The model does not satisfy all the conditions of the analytic solution:
the baryons are charged not neutral,
and they are not quite pressureless.
The baryons are charged for two reasons.
First, in the self-similar solutions
the charge of the black hole is created self-consistently by accretion,
so at least one stream must be charged.
Second,
mass inflation requires the simultaneous presence
of ingoing and outgoing streams.
If both streams are accreted from outside the black hole,
then one of the streams must be charged
in order that, repelled by the charge of the black hole,
the stream becomes outgoing inside the outer horizon.
The baryons are not quite pressureless,
having equation of state
$w = 10^{-6}$.
The reason for this is that the boundary conditions of the model
are set at a sonic point outside the black hole
where the infalling baryons accelerate from subsonic to supersonic.
For pressureless baryons,
$w = 0$,
the sonic point is at infinity.
Giving $w$ a small but finite value
brings the sonic point to a large but finite radius.

\partab

Figure~\ref{bu}
shows that, despite the fact that the model does not fulfill
all the assumptions of the analytic solution,
the fit is nevertheless very good (within the width of the line)
just before and during inflation.
Well before inflation,
the baryons depart from the fit because they are charged,
and therefore do not follow a geodesic.
The parameters of the fit are given in Table~\ref{par}.
The parameters
$C_a$
of the analytic solution~(\ref{betaami}),
namely
$C_b$ for the outgoing baryons
and $C_d$ for the ingoing dark matter,
are fits to the Reissner-Nordstr\"om phase,
equation~(\ref{betaa}),
entering inflation.
The values of the remaining parameters $\lambda$, $\mu_b$, and $\mu_d$
of the analytic solution~(\ref{betaami})
are given by equations~(\ref{lambda}) and (\ref{mub}),
measured from the model
at the onset of inflation, where $\beta$ goes through its minimum.
The parameters
$\lambda$, $\mu_b$, and $\mu_d$
are perhaps not what might have been naively expected
given the boundary conditions at the sonic point;
a discussion of why they
do in fact make sense
is relegated to Appendix~\ref{unexpected}.

\subsection{Collapse phase}
\label{collapsetwo}

During inflation, the streaming terms $\mu_b u_b^2$
on the right hand side of equation~(\ref{dlnbdlnutwo})
become huge, dwarfing the $\lambda$ terms.
But inflation also drives $\beta$ to large values,
eventually to the point where the $\beta^2$ mass terms
on the right hand side of equation~(\ref{dlnbdlnutwo})
can no longer be neglected compared to the
$\mu_b u_b^2$ streaming terms.
With the $\mu_b u_b^2$ and $\beta^2$ terms included,
equation~(\ref{dlnbdlnutwo}) becomes
\begin{equation}
\label{dlnbdlnuacollapse}
  {\dd \ln \beta \over \dd \ln u_a}
  =
  {
  \beta^2 + \mu_b u_b^2
  \over
  - \beta^2 + \mu_b u_b^2
  }
  \ .
\end{equation}
Equations~(\ref{dlnbdlnuacollapse})
constitute a pair of coupled ordinary differential equations,
and they have an exact solution given by the implicit equations
\begin{equation}
\label{ubacollapse}
  \beta u_a
  =
  {\lambda^2 \over C_a \mu_a}
  \left[
  {q_b + \mu_a u_a^2 / \beta^2
  \over
  q_b + \lambda^2 / (C_a^2 \mu_a)}
  \right]^{q_a/2}
  \ ,
\end{equation}
where $q_a$ are constants
\begin{equation}
  q_a \equiv
  {\lambda^2 + C_a^2 \mu_a
  \over
  C_a^2 \mu_a - C_b^2 \mu_b}
  \ ,
\end{equation}
and as before $a$ represents either stream, and $b$ the other.
Note that $q_a + q_b = 1$,
and that one of $q_a$ or $q_b$ is $< 0$ and the other is $> 1$.
For small accretion rates $\mu_a$ and $\mu_b$,
the constants $q_a$ and $q_b$ are large and of opposite sign,
so that the power law on the right hand side of equation~(\ref{ubacollapse})
approximates an exponential.
The overall factor in equation~(\ref{ubacollapse})
comes from matching to the inflationary solution~(\ref{betaamilate})
at $\beta^2 = \lambda$,
where
equation~(\ref{dlnbdlnuacollapse})
takes over from
equation~(\ref{dlnbdlnuami}).
Integration of equations~(\ref{dlnbudlnrtwo})
yields the radius $r_a$ of each stream:
\begin{equation}
\label{racollapse}
  r_a
  \propto
  \lvert u_a \rvert / \beta
  \ ,
\end{equation}
where the constant of proportionality may be obtained by matching
to equation~(\ref{rami})
with the inflationary solution~(\ref{betaamilate})
at $\beta^2 = \lambda$.

Equation~(\ref{dlnbdlnuacollapse})
shows that the velocity $u_a$ reaches a maximum
when $\beta^2 = \mu_b u_b^2$,
at which point equation~(\ref{ubacollapse}) shows that
\begin{equation}
\label{uamaxcollapse}
  u_a
  =
  {\lambda \, x_a^{1/4} \over C_a^{1/2} \mu_a^{3/4}}
  \left[
  {q_b + x_a
  \over
  q_b + \lambda^2 / (C_a^2 \mu_a)}
  \right]^{q_a/4}
  \ ,
\end{equation}
where $x_a \equiv - q_b / ( 1 + q_a )$
is the value of $\mu_a u_a^2 / \beta^2$ at the peak of $u_a$.

After the velocity peaks,
the matter collapses rapidly to zero radius.
The interior mass, equation~(\ref{M}), is
$M = \frac{1}{2} \beta^2 r_a$
to a good approximation,
which
according to equation~(\ref{racollapse})
is proportional to $\beta u_a$,
which
from equation~(\ref{ubacollapse})
tends to a constant at zero radius.
In the typical case of small $\mu_a$ and $\mu_b$,
the value of the interior mass $M$ attained at zero radius is
\begin{equation}
\label{Mcollapsetwo}
  M
  \approx
  {\lambda r_- \over 2}
  \exp
  \left[
  {\lambda^2 \ln ( C_a^2 \mu_a / C_b^2 \mu_b )
  \over
  2 ( C_a^2 \mu_a - C_b^2 \mu_b )}
  \right]
  \ ,
\end{equation}
which generalizes the equal stream result~(\ref{Mcollapse}).

\bucollapsefig

Figure~\ref{bucollapse}
compares the analytic solutions~(\ref{betaami}) and~(\ref{ubacollapse}),
valid respectively during the RN+inflation
and inflation+\discretionary{}{}{}collapse phases,
to a self-consistent self-similar model
whose parameters are listed in Table~\ref{par}.
Reaching the collapse phase in a model
requires a large accretion rate
$\mu$,
otherwise the peak velocity $u \sim \ee^{1/\mu}$
and related quantities overflow numerically.
A large accretion rate is more easily accomplished
in models where the baryons have pressure,
since then the sonic point is closer to the horizon.
To achieve the desired large accretion rate,
the model shown in Figure~\ref{bucollapse}
has an almost relativistic baryonic equation of state $w = 0.32$
(a value of $w$ slightly less than $1/3$
allows for an expected modest increase in the number of
relativistic particles species as temperature increases),
and a large dark-matter-to-baryon density
$\rho_d/\rho_b = 0.1$
at the sonic point.

Figure~\ref{bucollapse}
shows that the approximate solutions
provide a good fit to the behavior of
the neutral, pressureless dark matter,
and a satisfactory fit to the charged, relativistic baryons
from just before inflation up to the end of inflation.
As previously remarked for the model of Figure~\ref{bu},
the fit to the baryons fails well before inflation
because the baryons are charged, and therefore do not follow geodesics.
Figure~\ref{bucollapse}
also shows that the fit to the baryons fails
as soon as the collapse phase begins,
a fact that can be attributed to the pressure of the baryons,
since at this point the Lorentz force on the baryons is completely subdominant.
The parameters of the fit are listed in Table~\ref{par}.

The failure
in Figure~\ref{bucollapse}
of the fit to baryons in the collapse phase
is not terribly important.
What is important is that the analytic approximation~(\ref{ubacollapse})
captures reliably the behavior during inflation,
and it predicts approximately correctly when inflation comes to an end,
equation~(\ref{uamaxcollapse}).
Once collapse begins,
the ingoing and outgoing streams collapse to zero radius in a tiny instant.
Collapse to zero radius is subject to the non-radial
Belinski-Khalatnikov-Lifshitz
instabilities
\cite{Belinski:1970,Belinski:1971,Belinski:1982,Berger:2002st}.
Such instabilities,
which are artificially suppressed as long as spherical symmmetry is assumed,
and are not explored in this paper.

\buunbalancedfig

\subsection{Very unequal streams}
\label{unbalancedtwo}

What happens if one of the two streams
is very much smaller than the other?

Figure~\ref{buunbalanced}
illustrates a self-consistent self-similar model
in which the dark-matter-to-baryon density
has been set to a low value
$\rho_d / \rho_b = 10^{-6}$
at the sonic point.
Figure~\ref{buunbalanced}
shows that the smaller dark matter stream
undergoes pre-inflationary acceleration
that brings its velocity $u_d$
up to the point that
$C_d \mu_d u_d \approx C_b \mu_b u_b$,
in accordance with equation~(\ref{betaamilate}),
whereupon inflation ensues.
The parameters of the model, and of the fit to it,
are listed in Table~\ref{par}.

According to equation~(\ref{Mcollapsetwo}),
the maximum interior mass attained is of the order of
\begin{equation}
\label{Masym}
  M \sim \ee^{1/\mu_a}
\end{equation}
where $\mu_a$ is the {\em larger\/} of the two accretion rates.

\section{Baryon + dark matter models}
\label{baryondarkmatter}

The self-consistent models referred to in previous sections
are the general relativistic,
self-similar, accreting, spherical, charged black hole
solutions described by
\cite{Hamilton:2004av,Hamilton:2004aw},
to which the reader is referred for more detail.
In these models the black hole accretes two fluids:
charged ``baryons'' with equation of state $p_b / \rho_b = w$;
and neutral pressureless ``dark matter.''
The baryons, being repelled by the charge of the black hole
produced self-consistently from the accreted baryonic charge,
become outgoing inside the horizon.
The dark matter, being neutral, remains ingoing.
Relativistic counter-streaming between the baryons
and the dark matter produces mass inflation.


The boundary conditions of the self-similar solutions
are set outside the outer horizon,
at a sonic point where the infalling baryonic plasma accelerates
from subsonic to supersonic.
Setting boundary conditions outside the horizon is natural,
since information can propagate only inwards inside a black hole.
The sonic point is assumed to be regular, meaning that
the acceleration through the sonic point is finite and differentiable,
which sets two boundary conditions.
The accretion in real black holes is likely to be more complicated,
but this assumption is the simplest physically reasonable one.

Given the assumption of a regular sonic point,
the self-similar solutions are characterized by
four dimensionless free parameters:
(1) the charge-to-mass ratio $\Qbh / \Mbh$ of the black hole;
(2) the mass accretion rate $\Mbhdot$;
(3) the equation of state $w$ of the baryons;
and
(4) the dark-matter-to-baryon ratio $\rho_d / \rho_b$.
Being self-similar,
the solutions scale with black hole mass $\Mbh$,
which increases linearly with time.
\cite{Hamilton:2004av,Hamilton:2004aw} also consider models where the baryons have
a finite conductivity, and the dark matter has a finite
interaction cross-section with the baryons,
but the present paper assumes that such dissipative processes are not present.

\cite{Hamilton:2004av,Hamilton:2004aw} use a parameter $\eta_s$ in place of the
dimensionless mass accretion rate $\Mbhdot$;
the relation between the two is exhibited in Appendix~\ref{massaccretionrate}.

\darkmatteraccretionMfig

\darkmatteraccretionfig

\subsection{Dependence on accretion rate}
\label{darkmatteraccretionrate}

An important and surprising prediction of the analytic models
of
\S\S\ref{twoequalstreams} and \ref{twounequalstreams}
is that the smaller the accretion rate,
the larger the inflationary growth rate.
Specifically,
the inflationary growth rate $\dd \ln \beta / \dd \ln r_a$
[recall that the interior mass $M$ is related to $\beta$ by eq.~(\ref{M})]
of stream $a$
is inversely proportional to the ``accretion rate'' $\mu_b$
of the stream $b$,
equation~(\ref{dlnbetadlnrmitwo}),
where $a$ is either of the ingoing or outgoing streams,
and $b$ is the other stream.
This is a ferocious
kind of instability,
where the tiniest effect provokes the most violent reaction.

Figure~\ref{darkmatteraccretionM}
shows the interior mass $M$ as a function of radius
for self-similar models with three different mass accretion rates,
$\Mbhdot = 0.003$, $0.01$, and $0.03$,
demonstrating that indeed
the smaller the accretion rate, the faster the interior mass inflates.
The model with $\Mbhdot = 0.01$
is the same as that illustrated in Figure~\ref{bucollapse}.
The parameters of the models are listed in Table~\ref{par}.

Superposed on each model in
Figure~\ref{darkmatteraccretionM}
is the analytic prediction
from equation~(\ref{rami})
for the Reissner-Nordstr\"om+inflation phase,
and from equation~(\ref{racollapse})
for the inflation+collapse phase.
The good fits confirm that
the approximate analytic models developed
in \S\S\ref{twoequalstreams} and \ref{twounequalstreams}
are a reliable guide to inflationary behavior,
even though the models do not satisfy all the assumptions
of the analytic approximations:
the baryons are charged, not neutral,
and the equation of state of the baryons is relativistic, not pressureless.

\subsection{The streams remain cold}

Figure~\ref{darkmatteraccretion}
plots the energy density and Weyl curvature
for the same set of three models as Figure~\ref{darkmatteraccretionM}.
Figure~\ref{darkmatteraccretion}
shows that the Weyl curvature scalar $C$
and the center-of-mass energy density $\rho$
inflate like the interior mass $M$.

Figure~\ref{darkmatteraccretion}
also shows
the individual proper densities $\rho_b$ of baryons,
$\rho_d$ of dark matter, and $\rho_e$ of electromagnetic energy
for the model with
$\Mbhdot = 0.01$
(only this case is shown, to avoid confusion).
The individual densities remain modest
in spite of the fact that the center-of-mass energy density
is inflating exponentially.
This shows that almost all of the center-of-mass energy density
is in the streaming energy.
As in a particle accelerator,
the streams remain cold as they are accelerated through each other.

The fact that the proper densities of individual streams
remain modest during inflation
is consistent with the conclusion of \cite{Ori:1991}
that volume elements remain only mildly distorted during inflation,
even though the Weyl curvature exponentiates hugely.
The physical reason for the mild distortion
is that
only a tiny proper time goes by during inflation,
as discussed in \S\ref{future} below.
Although the tidal force becomes exponentially huge,
it does not have enough time to do damage.

\section{Single fluid models}
\label{singlefluid}

So far, all the models considered in this paper
have involved two separate fluids that stream through each other.
However,
the great majority of studies of mass inflation
inside spherical, charged black holes
have considered not two separate fluids,
but rather the case of a massless scalar field,
usually neutral
\cite{Christodoulou:1986a,Christodoulou:1986b,Christodoulou:1987a,Christodoulou:1987b,Goldwirth:1987,Gnedin:1993,Bonanno:1994qh,Brady:1995ni,Brady:1994aq,Burko:1997fc,Burko:1997zy,Burko:1998jz,Husain:2000vm,Burko:2002qr,Burko:2002fv,MartinGarcia:2003gz,Dafermos:2004jp,Hansen:2005am},
but sometimes charged
\cite{Hod:1996ar,Hod:1998gy,Hod:1999wb,Sorkin:2000pc,Oren:2003gp,Dafermos:2003wr,Dafermos:2003yw}.

A massless scalar field supports waves moving at the speed of light, so
it is plausible that mass inflation in a scalar field
arises physically from counter-streaming between ingoing and outgoing waves.
Indeed, this was \cite{Poisson:1990eh}'s original idea:
they recognized that relativistic counter-streaming
between ingoing and outgoing streams
would drive the instability that they termed mass inflation,
and they proposed that the outgoing stream
would be produced
by a Price tail of outgoing gravitational radiation
generated by gravitational collapse.
A massless scalar field is intended to model this radiation,
with the simplifying advantage that
a massless scalar field supports spherically symmetric waves.

As expounded by
\cite{Babichev:2008dy},
the equations governing a massless scalar field
are the same as those of a perfect fluid with
an ultra-hard equation of state,
$p = \rho$,
subject to the classical condition that
the gradient (momentum) of the field is everywhere timelike.
In fact the gradient of a scalar field is not necessarily
timelike everywhere,
so a scalar field is not completely equivalent to an ultra-hard perfect fluid.
However,
an ultra-hard fluid shares with a scalar field
the property that it supports waves moving at the speed of light.
In what follows
we suppose that the behavior of an ultra-hard fluid
may be an adequate guide to the behavior of a massless scalar field.

The models considered in this section are all self-similar,
belonging to the same suite of self-similar models considered in
\S\ref{baryondarkmatter},
but now for the case of a single fluid.

\subsection{Varying equation of state}
\label{varyeqstate}

\eqstatefig

Is it true that mass inflation in a massless scalar field
arises from relativistic counter-streaming between ingoing and outgoing waves?
Support for this idea comes from looking at what happens
in perfect fluids with various equations of state $w = p / \rho$.
The speed of sound in a perfect fluid is $\sqrt{w}$,
which equals the speed of light if $w = 1$,
but is less than the speed of light if $w < 1$.

The top panel of
Figure~\ref{eqstate}
shows the proper velocity $V$ of the similarity frame relative to the
infalling fluid for self-similar models with various equations of state $w$.
These are single-fluid models,
unlike what has been considered in previous sections.
The fluid is charged,
because in self-similar models the charge of the black hole
is produced self-consistently from the charge of the accreted fluid.
If the black hole is to be charged and hence to have an inner horizon
where mass inflation can take place,
then the accreted fluid must be charged.

The velocity $V$ in the top panel of
Figure~\ref{eqstate}
equals
\cite{Hamilton:2004av}
the ratio $\xi^r/\xi^t$
of the radial and time components of the homothetic 4-vector $\xi^m$
(the conformal Killing vector)
in the frame of the fluid.
The velocity $V$
essentially defines the horizon structure of self-similar black holes
(the fact that horizons can be defined unambiguously is associated with the
presence of conformal time translation symmetry, that is,
of self-similarity).
The velocity $V$ is subluminal outside horizons,
is equal to $\pm 1$ at horizons,
and is superluminal between horizons.
The sign of the velocity $V$
provides an alternative definition of what is meant by ingoing and outgoing:
positive $V$ is ingoing, while negative $V$ is outgoing.
This definition differs from the definition of ingoing and outgoing
adopted in the present paper,
which is based on the sign of the radial gradient $\beta_r$,
but the two definitions agree in the important region near the inner horizon.
The definition
adopted in the present paper
is to be preferred because it remains valid in arbitrary spherical spacetimes,
whereas the definition based on the sign of the velocity $V$ works
only if there is (conformal) time translation symmetry.
Figure~\ref{eqstate}
shows that the superluminal velocity $V$
goes through infinity from positive to negative
some way inside the horizon.
This signals that the rest frame of the fluid
changes from ingoing to outgoing,
which happens because the charged fluid is repelled
by the charge of the black hole.

Figure~\ref{eqstate}
shows that mass inflation begins in models with
sound speed near but less than the speed of light
($w$ near but less than $1$),
but then stalls.
Once inflation has stalled,
then the fluid drops through an outgoing inner horizon,
where the velocity is $V = -1$
(in the models with $w \ge 0.98$ in Fig.~\ref{eqstate},
the numerics overflow before the inner horizon is actually reached).
Of course, mass inflation would occur in these models
near the inner horizon
even if only the tiniest amount of ingoing stream were present;
but by assumption there is only a single fluid here.

As discussed in \S\ref{mechanism},
if ingoing and outgoing streams are simultaneously present at an inner horizon,
then necessarily the ingoing and outgoing streams
must be streaming through each other at the speed of light.
But if there is only a single fluid,
and the speed of sound in the fluid is less than the speed of light,
then that fluid cannot support both ingoing and outgoing waves
at the inner horizon.
If all streams are only ingoing, or only outgoing,
then the mechanism of mass inflation is thwarted,
and there is no obstacle to the streams dropping through an inner horizon,
as seen in Figure~\ref{eqstate}.

Figure~\ref{eqstate}
shows that only in the case where the speed of sound equals the speed of light,
$w = 1$,
does mass inflation persist.
The behavior illustrated by Figure~\ref{eqstate}
is thus consistent with the physical idea that
mass inflation in a massless scalar field
is driven by relativistic counter-streaming between
ingoing and outgoing waves.

\scalarfieldaccretionfig

\scalarmdotfig

\subsection{Dependence on accretion rate}
\label{scalaraccretionrate}

Figure~\ref{scalarfieldaccretion}
shows the
energy density $\rho_\phi$
and Weyl curvature scalar $\lvert C \rvert$
for self-similar models
accreting
an ultra-hard fluid,
$w \equiv p_\phi / \rho_\phi = 1$,
at three different mass accretion rates
$\Mbhdot = 0.001$, $0.003$, and $0.01$.
The model with $\Mbhdot = 0.01$ was
already featured in Figure~\ref{eqstate}.

Figure~\ref{scalarfieldaccretion}
shows that the inflationary growth rate
is faster for smaller accretion rates,
in agreement with expectations from the analytic models of
\S\S\ref{twoequalstreams} and \ref{twounequalstreams}.
As illustrated in Figure~\ref{scalarmdot},
the inflationary growth rate
is inversely proportional to the mass accretion rate,
except at the largest accretion rates.
For accretion rates away from the maximum,
the curve shown in Figure~\ref{scalarmdot} fits to
\begin{equation}
\label{scalardlnbdlnr}
  {\dd \ln \beta \over \dd \ln r}
  \approx
  -
  {0.47 \over \Mbhdot}
  \ .
\end{equation}
The maximum possible accretion rate
is constrained
by the requirement of appropriate boundary conditions at the sonic point
\cite{Hamilton:2004av}.
For
an ultra-hard fluid
accreting on to a black hole of charge-to-mass $\Qbh/\Mbh = 0.8$,
the maximum accretion rate is
$\Mbhdot = 0.0784$.

Figure~\ref{scalarfieldaccretion} shows that
(at least for constant accretion rate,
as is necessarily true in self-similar models)
the
ultra-hard fluid
shows a single power-law inflationary phase,
as opposed to
two distinct phases of inflation followed by collapse.
The
ultra-hard fluid
does collapse to zero radius
(if quantum gravity does not intervene),
but not as a separate phase.
The interior mass tends to infinity at zero radius,
as opposed to reaching a finite value.

A prominent difference between the
ultra-hard fluid
and the two-stream models considered in previous sections
is that the
ultra-hard fluid
always maintains a large transverse pressure.
Of course this is built in to the equation of state
$p = \rho$
of the
ultra-hard fluid.
But it does contrast with the two-stream models,
where inflation accelerates ingoing and outgoing streams in opposite directions,
which produces an exponentially growing radial pressure
in the center-of-mass frame,
but no transverse pressure to go with it.
Which is a correct, a large or a small transverse pressure?
The authors' inclination is to suspect
that the two-stream model, with small transverse pressure,
is likely to be a closer model of reality,
on the grounds that
the feedback loop that drives inflation
involves only radial accelerations
and radial components of the energy-momentum tensor,
with no dependence on transverse pressure.
Absent some physical mechanism to maintain isotropy,
it is hard to see how the transverse pressure could keep up
with the exponentially growing radial pressure.

\section{The far future?}
\label{future}

The Penrose diagram of a Reissner-Nordstr\"om
or Kerr-Newman black hole
indicates that an observer who passes through the outgoing inner horizon
sees the entire future of the outside universe go by.
In a sense,
this is ``why'' the outside universe appears infinitely blueshifted.

This raises the question of whether what happens
at the outgoing inner horizon of a real black hole
indeed depends on what happens in the far future.
If it did,
then the conclusions of previous sections,
which are based in part
on the proposition that the accretion rate is approximately constant,
would be suspect.
A lot can happen in the far future,
such as black hole mergers,
the Universe ending in a big crunch,
Hawking evaporation,
or something else beyond our current ken.

Ingoing and outgoing observers
both see each other highly blueshifted near the inner horizon.
An outgoing observer sees ingoing observers accreted
to the future of the time that the outgoing observer fell in,
while an ingoing observer sees outgoing observers accreted
to the past of when the ingoing observer fell in.
If the streaming 4-velocity between ingoing and outgoing
streams is $u_{ba}^m$,
equation~(\ref{uba}),
then
the proper time $\dd \tau_b$ that elapses on stream $b$
observed by stream $a$
equals the blueshift factor $u_{ba}^t$
times the proper time
$\dd \tau_a$ experienced by stream $a$,
\begin{equation}
  \dd \tau_b
  =
  u_{ba}^t \,
  \dd \tau_a
  \ .
\end{equation}

\subsection{Inflationary phase}
\label{inflationfuture}

A physically relevant timescale for the observing stream $a$
is how long it takes for the blueshift to increase by one $\ee$-fold,
which is
${\dd \tau_a / \dd \ln u_{ba}^t}$.
During the inflationary phase,
stream $a$ sees the amount of time elapsed on stream $b$
through one $\ee$-fold of blueshift to be
\begin{equation}
\label{dtbmi}
  u_{ba}^t \,
  {\dd \tau_a \over \dd \ln u_{ba}^t}
  =
  {2 C_b r_- \over \lambda}
  \ .
\end{equation}
The right hand side of equation~(\ref{dtbmi}) is derived from
$u_{ba}^t \approx 2 \lvert u_a u_b \rvert$,
$\lvert \dd r_a / \dd \tau_a \rvert = \lvert \beta_{a, t} \rvert \approx \beta u_a$,
$r_a \approx r_-$,
and the approximations~(\ref{betaamilate})
and (\ref{dlnbetadlnrmitwo})
valid during the inflationary phase.
The constants $C_b$ and $\lambda$
on the right hand side of
equation~(\ref{dtbmi})
are typically of order unity,
while $r_-$ is the radius of the inner horizon where mass inflation takes place.
Thus the right hand side of
equation~(\ref{dtbmi})
is of the order of one black hole crossing time.
In other words,
stream $a$ sees
approximately one black hole crossing time
elapse on stream $b$
for each $\ee$-fold of blueshift.
The estimate~(\ref{dtbmi}) is compatible with formula~(1) of \cite{Burko:1996sm},
who considered an outgoing infaller irradiated by cosmic microwave background
photons, truncated when the blueshifted photons reached Planck energy.

For astronomically realistic black holes,
exponentiating the Weyl curvature up to the Planck scale
will take typically a few hundred $\ee$-folds of blueshift,
as illustrated for example in Figure~\ref{darkmatteraccretion}.
Thus what happens at the inner horizon of a realistic black hole
before quantum gravity intervenes
depends only on the immediate past and future of the black hole
--
a few hundred black hole crossing times
--
not on the distant future or past.
This conclusion holds even if
the accretion rate of one of the ingoing or outgoing streams
is tiny compared to the other,
as considered in \S\ref{unbalancedtwo}.

From a stream's own point of view, the entire inflationary episode
goes by in a flash.
At the onset of inflation,
where $\beta$ goes through its minimum,
at
$\mu_a u_a^2 = \mu_b u_b^2 = \lambda$
according to equation~(\ref{dlnbdlnuami}),
the blueshift
$u_{ba}^t$
is already large
\begin{equation}
  u_{ba}^t
  =
  {2 \lambda \over \sqrt{\mu_a \mu_b}}
  \ ,
\end{equation}
thanks to the small accretion rates $\mu_a$ and $\mu_b$.
During the first $\ee$-fold of blueshift,
each stream experiences a proper time of order
$\sqrt{\mu_a \mu_b}$ times the black hole crossing time,
which is tiny.
Subsequent $\ee$-folds of blueshift race by
in proportionately shorter proper times.
The time is so short that volume elements of each stream
are little distorted during inflation despite the huge tidal force,
as first pointed out by \cite{Ori:1991}.

\subsection{Collapse phase}
\label{collapsefuture}

The time to reach the collapse phase is another matter.
According to the estimates in \S\S\ref{collapsetwo}
and \ref{unbalancedtwo},
reaching the collapse phase takes
of order $\sim 1/\mu_a$ $\ee$-folds of blueshift,
where $\mu_a$ is the larger of the accretion rates of
the ingoing and outgoing streams,
equation~(\ref{Masym}).
Thus in reaching the collapse phase,
each stream has seen approximately $1/\mu_a$
black hole crossing times elapse on the other stream.
But $1/\mu_a$ black hole crossing times
is just the accretion time
--
essentially, how long the black hole has existed.
This timescale, the age of the black hole, is not infinite,
but it can hardly be expected that the accretion rate of a
black hole would be constant over its lifetime.

If the accretion rate were in fact constant,
and if quantum gravity did not intervene
and the streams remained non-interacting,
then indeed streams inside the black hole would reach the collapse phase,
whereupon they would plunge to a spacelike singularity at zero radius.
For example, in the self-similar models
illustrated in Figures~\ref{darkmatteraccretionM} and \ref{darkmatteraccretion},
outgoing baryons hitting the central singularity
see ingoing dark matter accreted a factor of two into the future
(specifically,
for ingoing baryons and outgoing dark matter hitting the singularity,
the radius of the outer horizon when the dark matter is accreted
is twice that when the baryons were accreted;
the numbers are $2.11$ for $\Mbhdot = 0.03$,
$1.97$ for $\Mbhdot = 0.01$,
and unknown for $\Mbhdot = 0.003$ because in that case
the numerics overflow before the central singularity is reached).
The same conclusion applies to the
ultra-hard fluid
models
illustrated in Figure~\ref{scalarfieldaccretion}:
if the accretion rate is constant,
then outgoing streams see only a factor of order unity or a few into the future
before hitting the central singularity.

If on the other hard the accretion rate decreases
sufficiently rapidly with time,
then it is possible
that an outgoing stream never reaches the collapse phase,
because the number of $\ee$-folds to reach the collapse phase
just keeps increasing as the accretion rate decreases.
By contrast,
an ingoing stream is always liable to reach the collapse phase,
if quantum gravity does not intervene,
because an ingoing stream sees the outgoing stream from the past,
when the accretion rate was liable to have been larger.

However,
as already remarked in \S\ref{collapsesym},
in astronomically realistic black holes,
it is only for large accretion rates,
such as may occur when the black hole first collapses
($\Mbhdot \gtrsim 0.01$
for the models illustrated in Fig.~\ref{darkmatteraccretion}),
that the collapse phase has a chance of being reached before the Weyl curvature
exceeds the Planck scale.

To summarize,
it is only streams accreted during the first few hundred or so
black hole crossing times
following a black hole's formation,
or following an exceptionally high accretion event such as a black hole merger,
that have a chance of hitting a central spacelike singularity.
Streams accreted at later times,
whether ingoing or outgoing,
are likely to meet their fate
in the inflationary zone at the inner horizon,
where the Weyl curvature exponentiates to the Planck scale and beyond.

\subsection{Null singularity on the Cauchy horizon?}
\label{nullsingularity}

\cite{Dafermos:2003wr}
has proved
a number of mathematical theorems
that establish that
a null singularity forms on the Cauchy horizon
of a charged spherical black hole
accreting a massless scalar field.
The situation envisaged by the theorems is that of a black hole
that collapses and thereafter remains isolated.
The collapse generates an outgoing Price tail of radiation.
The theorems assume that the outgoing Price radiation
falls off sufficiently rapidly along outgoing null geodesics,
and
\cite{Dafermos:2003yw}
have proved that the required condition on the Price radiation holds
for an isolated spherical black hole accreting a massless scalar field.
The theorems confirm the several analytic and numerical studies
that have found a null singularity on the Cauchy horizon
\cite{Ori:1991,Bonanno:1994qh,Brady:1995ni,Burko:1997zy,Burko:1997fc,Hod:1998gy,Hod:1999wb,Ori:2001pc,Hansen:2005am}.

The conclusion is completely consistent with
the results of the present paper.
However,
as regards real astronomical black holes,
the question of whether a null singularity forms is academic.
As emphasized by \cite{Poisson:1990eh}
and discussed in 
\S\ref{inflationfuture},
quantum gravity is likely to intervene.
If quantum gravity is set aside,
then, as discussed in \S\ref{collapsefuture},
the appearance of a null singularity depends on events happening
in the indefinite future.

\cite{Burko:2002qr,Burko:2002fv}
find numerically that a null singularity forms only if the scalar field
set up outside the horizon falls off sufficiently rapidly,
the required degree of rapidity depending on the parameters of the problem,
such as the charge-to-mass ratio of the black hole.
If too much scalar field continues to be accreted,
then no null singularity forms,
and the field collapses to a central singularity.
These results are consistent with the arguments of the present paper.
Again however, we emphasize that the appearance of a null singularity
depends on events happening in the indefinite future,
and neglects quantum effects that undoubtedly will be important,
whether those arise from super-Planckian densities and curvatures,
or from super-Planckian collisions between ingoing and outgoing streams,
or from Hawking evaporation of the black hole,
or from pair creation near the inner horizon.


\cite{Frolov:2006is}
have shown that in the simplified case of
a 1+1-dimensional charged black hole,
if the effects of pair creation of charged particles are taken into account,
then the result is collapse to a spacelike singularity
rather than a null singularity on the Cauchy horizon.
The result is consistent with the argument of the present paper
that as long as there is any source that continues
to replenish ingoing and outgoing streams
near the inner horizon, the ultimate result will be collapse
to a spacelike singularity.
The results of
\cite{Frolov:2006is}
suggest that even without any direct accretion,
pair creation provides a sufficient source of ingoing and outgoing streams.

\section{Collision rate}
\label{collisionrate}

It has been assumed throughout this paper that ingoing and outgoing streams
stream through each other without interacting.
The aim of this section is to check the validity of that assumption.
The conclusion is that the assumption is good
at typical astronomically low accretion rates,
but is likely to break down at higher accretion rates.
Moreover the assumption may well break down at center-of-mass
collision energies exceeding the Planck mass.
We hope to explore the consequences of interaction between streams
in a subsequent paper.

For definiteness,
the models considered in this section are the baryon-plus-dark-matter models
illustrated in Figure~\ref{darkmatteraccretion}.
As a practical matter, it is unlikely that,
as assumed by the models,
the ingoing stream would consist
only of dark matter and the outgoing stream only of baryons.
More realistically, both streams would contain baryons,
and baryon-baryon collisions would dominate baryon-dark matter collisions,
at least at low energies.
However,
it is convenient to refer to the two streams
as ``baryon'' and ``dark matter'' streams.

\bhcolliderfig

Each baryon in the black hole particle accelerator
sees a flux
$n_d u$
of dark matter particles per unit area per unit time,
where $n_d = \rho_d / m_d$
is the proper number density of dark matter particles in their own frame,
and $u \equiv u_{db}^r$
is the radial component of the proper streaming 4-velocity,
the $\gamma v$,
of the dark matter through the baryons
[for brevity,
in this subsection
the streaming velocity $u_{db}^r$,
equation~(\ref{uba}),
is written simply as $u$;
this is not the same as the $u$
defined by equation~(\ref{betamu})].
The $\gamma$ factor in $u$ is the relativistic beaming factor:
all frequencies, including the collision frequency,
are speeded up by the relativistic beaming factor $\gamma$.
As the baryons accelerate through the collider,
they spend a proper time interval $\dd \tau / \dd \ln u$
in each $\ee$-fold of velocity $u$.
The number of collisions per baryon per $\ee$-fold of $u$
is the dark matter flux $( \rho_d / m_d ) u$,
multiplied by the time
$\dd \tau / \dd \ln u$,
multiplied by the collision cross-section $\sigma$.
\begin{equation}
\label{collisionsb}
  {\mbox{number of collisions}
  \over
  \mbox{baryon $\times$ $\ee$-fold of $u$}}
  \approx
  {\rho_d \over m_d} \sigma u
  {\dd \tau \over \dd \ln u}
  \ .
\end{equation}
During the inflationary phase,
the collision rate coefficient
$\rho_d u \, {\dd \tau / \dd \ln u}$
that goes into equation~(\ref{collisionsb}) is,
from equation~(\ref{dtbmi})
and the definition~(\ref{mub}) of $\mu_d$,
\begin{equation}
  \rho_d u
  {\dd \tau \over \dd \ln u}
  =
  {C_d \mu_d \over 8 \pi \lambda r_-}
  \propto
  {\Mbhdot \over \Mbh}
  \ ,
\end{equation}
in which the proportionality on the rightmost side
follows from $\mu_d \propto \Mbhdot$
and $r_- \propto \Mbh$.
Figure~\ref{bhcollider}
shows, for several different mass accretion rates $\Mbhdot$,
the collision rate
$\rho_d u \, \dd \tau / \dd \ln u$
of the black hole collider,
expressed in units of the inverse black hole accretion time
$\Mbhdot / \Mbh$.
In the units $c = G = 1$ being used here,
the mass of a baryon (proton) is
$1 \unit{GeV} \approx 10^{-54} \unit{m}$.
If the cross-section $\sigma$ is expressed in units of femtobarns
($1 \unit{fb} = 10^{-43} \unit{m}^2$),
which is about a weak interaction cross-section,
then the number of collisions~(\ref{collisionsb})
per baryon per $\ee$-fold of collision velocity $u$ is
\begin{align}
\label{collisionsbn}
  &
  {\mbox{number of collisions}
  \over
  \mbox{baryon $\times$ $\ee$-fold of $u$}}
  \approx
  10^{-19}
  \left( {\sigma \over 1 \unit{fb}} \right)
  \left( {300 {\unit{GeV}} \over m_d} \right)
\nn
  &
  \qquad\qquad
  \times \,
  \left( {\Mbhdot / \Mbh \over (10^{10} \unit{yr})^{-1}} \right)
  \left( {\rho_d u \, \dd \tau / \dd \ln u \over 0.03 \, \Mbhdot / \Mbh} \right)
  \ .
\end{align}
The number~(\ref{collisionsbn})
of collisions per baryon is small.
This is like a particle accelerator,
where collisions are rare.
If the dark matter stream is dominated by baryons,
as may well be true in realistic models,
then the collision rate~(\ref{collisionsbn})
rises to of order unity for low-momentum-transfer
electromagnetic or strong collisions,
where cross-sections approach one barn.
However,
a few low-momentum-transfer collisions
will not spoil the conclusion that to a good approximation
the counter-streaming streams are non-interacting,
at least in the realm of ``known physics,''
at sub-Planckian center-of-mass collision energies.

The collision rate~(\ref{collisionsbn})
is scaled to the typical accretion timescale
$\Mbh / \Mbhdot \sim 10^{10} \unit{yr}$
of astronomical black holes.
The collision rate would be higher during episodes of high accretion.
At sufficiently high accretion rates,
collisions between streams would affect the streams,
and the assumption of non-interacting streams would break down.


Stars in Figure~\ref{bhcollider}
mark where
the center-of-mass energy
$\sqrt{m_b m_d u}$
of colliding baryons and dark matter particles,
taken to have masses $m_b = 1 \unit{GeV}$ and $m_d = 300 \unit{GeV}$,
hits the Planck energy.
This occurs well before the curvature hits the Planck scale.
If cross-sections increase sufficiently rapidly
at super-Planckian center-of-mass energies,
then again collisions between streams would become important,
and the assumption of non-interacting streams would fail.


The total number of collisions taking place in the black hole
per unit (external) time
equals the rate~(\ref{collisionsbn}) per baryon
multiplied by rate at which the black hole is accreting baryons,
which is approximately equal to the mass accretion rate $\Mbhdot$
divided by the mass $m_b$ per baryon:
\begin{equation}
\label{collisions}
  {\mbox{number of collisions}
  \over
  \mbox{time $\times$ $\ee$-fold of $u$}}
  \approx
  {\Mbhdot \over m_b} {\rho_d \over m_d} \sigma u
  {\dd \tau \over \dd \ln u}
  \ .
\end{equation}
Numerically,
the number of collisions~(\ref{collisions}) per unit (external) time is
\begin{align}
\label{collisionsM}
  &
  {\mbox{number of collisions}
  \over
  \mbox{time $\times$ $\ee$-fold of $u$}}
  \approx
  10^{35}
  {\unit{yr}}^{-1}
  \left( {\sigma \over 1 \unit{fb}} \right)
  \left( {300 {\unit{GeV}}^2 \over m_b m_d} \right)
\nn
  &
  \quad
  \times \,
  \left( {\Mbhdot \over 10^{-16}} \right)
  \left( {\Mbhdot / \Mbh \over (10^{10} \unit{yr})^{-1}} \right)
  \left( {\rho_d u \, \dd \tau / \dd \ln u \over 0.03 \, \Mbhdot / \Mbh} \right)
  \ .
\end{align}
The collision rate~(\ref{collisionsM})
is scaled to the accretion rate of the
Milky Way black hole, \S\ref{sgra}.
Current-era particle accelerators
are proud of delivering inverse femtobarns of luminosity
during their lives.
Equation~(\ref{collisionsM})
shows that the accelerator
inside the Milky Way black hole
is delivering of order $10^{35}$ inverse femtobarns per year
in each $\ee$-fold of collision velocity
up to the Planck energy and beyond.

\section{Summary}
\label{summary}

The purpose of this paper has been to give a clear
account of the physical causes underlying the mass inflation instability
first proposed by \cite{Poisson:1990eh}.
The arguments have been restricted to the case of spherical,
charged black holes,
which has been the case considered by most studies of mass inflation to date.
It is generally thought that charge may be a satisfactory
surrogate for spin,
since charged black holes have inner horizons like rotating black holes.

The elements of mass inflation
can be seen already in the two Einstein equations~(\ref{dtbspherical}),
valid for arbitrary spherically symmetric spacetimes.
The first~(\ref{dtbtspherical}) of these equations relates the acceleration
to the gravitational force,
as measured by a freely-falling observer.
Acceleration here means
the proper rate of change
of the proper radial velocity $\beta_t \equiv \partial_t r$
measured by the observer.
The equation~(\ref{dtbtspherical})
differs from the corresponding Newtonian equation
in that the gravitational force is sourced not only by the interior mass,
but also by radial pressure.
A second essential difference compared to Newtonian gravity
is that gravity in spherical spacetimes is governed by two equations, not one.
The radial velocity $\beta_t$
is one component of a 4-vector,
the radial 4-gradient
$\beta_m \equiv \partial_m r \equiv \{ \beta_t , \beta_r , 0 , 0 \}$.
The Einstein equation~(\ref{dtbrspherical})
for the second component $\beta_r$
is also central to producing mass inflation.

As originally argued by \cite{Poisson:1990eh},
mass inflation is produced by relativistic counter-streaming
between ingoing and outgoing streams just above the inner horizon.
The present paper has shown that the counter-streaming
is driven by the gravitational force
produced by the streaming pressure and energy flux.
This is why mass inflation is exponential:
the streaming pressure and flux
increase the gravitational force,
which accelerates the streams faster through each other,
which increases the streaming pressure and flux,
which increases the gravitational force,
and so on.

A crucial ingredient of the process is that
the gravitational force acts in opposite directions
for ingoing and outgoing streams:
it accelerates ingoing streams towards the black hole,
and outgoing streams away from the black hole.
\cite{Hamilton:2004aw}
explain this conundrum by pointing out that the gravitational
force is always inwards, in the direction of smaller radius,
and that the direction of smaller radius
is towards the black hole for ingoing streams,
and away from the black hole for outgoing streams.
The Einstein equation~(\ref{dtbrspherical})
shows that this picture is essentially correct,
although as argued in the next paragraph it is too simplistic.
A stream is ingoing or outgoing depending on whether
the proper radial gradient $\beta_r \equiv \partial_r r$
measured in that frame is positive or negative,
and the Einstein equation~(\ref{dtbrspherical})
shows that $\beta_r$ increases or decreases
depending on whether the energy flux in that frame
is positive or negative.
The result is that gravity tends to drive
a positive (ingoing) $\beta_r$ to become more positive,
and a negative (outgoing) $\beta_r$ to become more negative.

However,
this picture that the gravitational force points
in the direction of smaller radius,
which is in opposite directions for ingoing and outgoing streams,
does not explain why mass inflation can come to an end,
as found in the simple analytic models constructed
in \S\S\ref{twoequalstreams} and \ref{twounequalstreams}.
The more complete picture painted by the
Einstein equations~(\ref{dtbspherical})
is that the 4-vector nature of $\beta_m$,
whose time component is the radial velocity $\beta_t$, is central.
The ingoing and outgoing streams are accelerated through each other
only if the change in the radial component $\beta_r$
exceeds the change in the time component $\beta_t$,
as illustrated in the middle panel of Figure~\ref{betadiagram}.
If the change in $\beta_r$ is less than that in $\beta_t$,
then the streaming velocity decreases,
as illustrated in the bottom panel of Figure~\ref{betadiagram}.
If this happens,
then mass inflation comes to an end.

Sections~\ref{twoequalstreams} and \ref{twounequalstreams}
build on the qualitative picture of \S\ref{mechanism}
to develop simple approximate analytic models that capture
quantitatively how inflation is ignited, exponentiates, and then ends.
The first of the two sections, \S\ref{twoequalstreams},
considers the simple but unrealistic example
of symmetrically equal ingoing and outgoing streams,
which
\cite{Burko:1997xa,Burko:1998az,Burko:1998jz}
has previously called the homogeneous approximation
(see \S\ref{stationaryapproximation}).
The second of the two sections, \S\ref{twounequalstreams},
considers the realistic case of unequal ingoing and outgoing streams.
The reader interested in understanding mathematically
how Einstein's equations~(\ref{dtbspherical}) imply mass inflation
is encouraged to try the problem in Appendix~\ref{problem}.

The predictions of the analytic approximations are borne out
by comparison to self-consistent self-similar solutions.
Section~\ref{baryondarkmatter}
considers self-similar solutions in which the black hole
accretes two separate streams, ``baryons'' and ``dark matter,''
while \S\ref{singlefluid}
considers self-similar solutions in which the black hole accretes
just a single fluid.
The single-fluid models include the case of
a perfect fluid with an ultra-hard equation of state
$p = \rho$,
which can be taken as a model of a massless scalar field,
subject to the constraint that the momentum of the field
is everywhere timelike
\cite{Babichev:2008dy}.

In the single-fluid models considered in \S\ref{singlefluid},
it is shown that mass inflation occurs only in the case
of the ultra-hard fluid,
whose sound speed equals the speed of light.
If the equation of state $w = p / \rho$ is near but not equal to $1$,
so that the sound speed $\sqrt{w}$ is near but not
equal to the speed of light,
then mass inflation begins, but then stalls,
whereupon the outgoing charged fluid drops through the outgoing inner horizon.
The behavior is consistent with
the notion that mass inflation in
a fluid that supports waves moving at the speed of light,
such as a massless scalar field,
results from relativistic counter-streaming between ingoing and outgoing waves,
as originally suggested by \cite{Poisson:1990eh}.

Two possibly surprising predictions of the analytic solutions are:
first, that the smaller the accretion rate,
the faster inflation grows;
and second, that inflation eventually comes to an end,
whereupon the ingoing and outgoing streams collapse
to a spacelike singularity at zero radius.
Both predictions are confirmed in the self-consistent solutions,
in both two-fluid and single-fluid
cases.

The fact that a smaller accretion rate causes faster inflation
means that the inflationary instability is difficult to avoid.
How is the beast to be tamed
if the tiniest effect provokes the most violent reaction?

One possible way to avoid inflation is to introduce
a large amount of dissipation \cite{Wallace:2008zz},
fast enough to transport charge and angular momentum
so that the black hole becomes neutral and non-spinning
towards its center,
eliminating the inner horizon where inflation is ignited.
Given that angular momentum transport is a rather weak process
\cite{Balbus:1998},
we are inclined to suspect that real rotating black holes
do not dissipate all their spin,
and that inflation does occur in reality.

The results of this paper are consistent with all
the previous literature on the mass inflation instability.
It might seem that there is a conflict
in that we do not confirm the often-stated conclusion
that the generic consequence of inflation is
a weak null singularity on the Cauchy horizon.
The apparent conflict arises because
most studies have considered the situation of a black hole
that collapses and thereafter remains isolated,
whereas we have taken the point of view that
a real black hole is never isolated.
As long as a black hole continues to accrete,
and thereby to generate ingoing and outgoing streams
near the inner horizon,
the inflationary growth rate remains finite,
and there is no null singularity.
Even in the absence of direct accretion,
quantum-mechanical pair creation
provides a source of ingoing and outgoing streams
\cite{Frolov:2006is},
preventing a null singularity.

For typical astronomically low accretion rates,
inflation drives the Weyl curvature and
the center-of-mass energy density and pressure
far above the Planck scale
before significant tidal distortion occurs,
as first pointed out by
\cite{Ori:1991}.
Physically, even though tidal forces grow super-Planckian,
the timescale for those forces to act is so short
that volume elements remain little distorted.
Consequently,
as already pointed out by \cite{Poisson:1990eh},
quantum gravity in some form is likely to intervene before any
classical general relativistic singularity is reached.

If on the other hand the accretion rate is large enough,
as happens when the black hole first collapses,
and during events of high accretion such as a black hole merger,
then the streams collapse to a central singularity
before inflation drives the curvature above the Planck scale.
Such a crushing singularity is subject to non-radial
Belinski-Khalatnikov-Lifshitz
instabilities
\cite{Belinski:1970,Belinski:1971,Belinski:1982},
which have not been explored in this paper;
see \cite{Berger:2002st} for a review.

\section{Final remarks}
\label{finalremarks}

\subsection{Misconceptions}

As a pedagogical device,
it is helpful to confront a number of misconceptions
about what happens inside black holes.
The following remarks are predicated on the assumption
that what happens in a spherical, charged black hole,
as considered in this paper,
is a reliable guide to what happens in a realistic rotating black hole.

Misconception 1.
A popular story is that
matter that falls inside a black hole
falls to a central singularity,
a point of infinite curvature,
where space and time stop.
This misconception is based on the prototype of the Schwarzschild geometry,
bolstered by a simplistic interpretation
of the Penrose-Hawking
\cite{Penrose:1965,Hawking:1975}
singularity theorems.
In reality, as long as a black hole has any spin,
as is certain to be true
(the converse is a set of measure zero),
and except during epochs of high accretion,
such as within a few hundred black hole crossing times
of the initial collapse or of a black hole merger,
then matter falling into the black hole will reach its nemesis
in the inflationary region at the inner horizon,
which will be at a macroscopic radius
(some fraction of the radius of the outer horizon)
as long as the spin of the black hole is appreciably different from zero.
Except during epochs of high accretion,
inflation exponentiates the curvature and proper density
beyond the Planck scale
before collapse to a central singularity occurs.
What happens after the curvature and density have reached Planck scale
depends on quantum gravitational
processes that have yet to be explored in the literature.

Misconception 2
(\S\ref{future}).
What happens inside a black hole depends on what happens in the infinite future.
This misconception comes from looking at the Penrose diagram
of the Reissner-Nordstr\"om or Kerr-Newman geometry.
In reality,
during inflation,
an outgoing observer sees approximately one black hole crossing time
elapse on ingoing matter for each $\ee$-fold of blueshift
(an outgoing observer sees ingoing matter accreted from the future,
while an ingoing observer sees outgoing matter accreted in the past).
After a few hundred $\ee$-folds, the curvature will have exponentiated
to the Planck scale, whereupon quantum gravity presumably intervenes.
Far from depending on the infinite future,
what happens to an outgoing or ingoing stream
undergoing inflation depends on events
only a few hundred black hole crossing times
into the future or past.

Misconception 3
(\S\ref{nullsingularity}).
Mass inflation produces a null singularity on the Cauchy horizon.
As proven by 
\cite{Dafermos:2003wr},
confirming earlier studies
\cite{Ori:1991,Bonanno:1994qh,Brady:1995ni,Burko:1997zy,Burko:1997fc,Hod:1998gy,Hod:1999wb,Ori:2001pc},
a null singularity occurs on the Cauchy horizon
(outgoing inner horizon)
of a black hole that collapses and thereafter remains isolated for ever.
However,
a real black hole is never isolated,
and for ever is a long time.
It is true that an (outgoing or ingoing) observer who falls into
an astronomical black hole
is likely to meet their fate at the inner horizon
where inflation exponentiates the curvature to the Planck scale and beyond.
But that fate is something to do with quantum gravity,
not a general relativistic null singularity.

\subsection{The black hole particle accelerator}

We conclude this paper on a note of wonder.
If the story told in this paper is true,
then Nature has devised a most remarkable accelerator
of extraordinary power inside black holes.
The accelerator is powered by gravity, and it feeds on itself:
the higher the streaming energy,
the stronger the gravitational acceleration.
The accelerator is ingeniously constructed so that the smaller
the initial streams,
the more rapidly the acceleration exponentiates,
ensuring that nothing can escape it.
What does Nature do with such a machine?

\section*{Acknowledgements}
This work was supported in part by NSF award
AST-0708607.
PPA appreciates a Visiting Fellowship at JILA,
where this work was initiated.


\bibliographystyle{unsrt}
\bibliography{bh}

\appendix

\section{Spherically symmetric spacetime}
\label{sphericalspacetime}

The orthonormal tetrad formalism
introduces at each point of spacetime a tetrad,
a set of orthonormal axes
$\bgamma_m$.
The tetrad is related
to the basis $\bg_\mu$ of coordinate tangent vectors
by the vierbein
$e_m{}^\mu$
and its inverse
$e^m{}_\mu$
\begin{equation}
  \bgamma_m
  =
  e_m{}^\mu \bg_\mu
  \ , \quad
  \bg_\mu
  =
  e^m{}_\mu \bgamma_m
  \ .
\end{equation}
By construction,
the scalar products of the orthonormal tetrad axes $\bgamma_m$
constitute the Minkowski metric $\eta_{mn}$,
while by definition, the scalar products of the tangent basis $\bg_\mu$
constitute the coordinate metric $g_{\mu\nu}$
\begin{equation}
  \bgamma_m \cdot \bgamma_n
  \equiv
  \eta_{mn}
  \ , \quad
  \bg_\mu \cdot \bg_\nu
  \equiv
  g_{\mu\nu}
  \ .
\end{equation}
The line-element~(\ref{metric}) encodes not only a metric
but also an inverse vierbein
$e^m{}_\mu$, through
\begin{subequations}
\begin{align}
  e^t{}_\mu \, \dd x^\mu
  &=
  dt / \alpha
  \ ,
\\
  e^r{}_\mu \, \dd x^\mu
  &=
  ( 1 / \beta_r )
  ( \dd r - \beta_t \, \dd t / \alpha )
  \ ,
\\
  e^\theta{}_\mu \, \dd x^\mu
  &=
  r \, \dd \theta
  \ ,
\\
  e^\phi{}_\mu \, \dd x^\mu
  &=
  r \sin\theta \, \dd \phi
  \ .
\end{align}
\end{subequations}
Explicitly, the inverse vierbein
$e^m{}_\mu$ is
\begin{equation}
\label{inversevierbein}
  e^m{}_\mu
  =
  \left(
  \begin{array}{cccc}
  1 / \alpha & 0 & 0 & 0 \\
  - \beta_t / ( \alpha \beta_r ) & 1 / \beta_r & 0 & 0 \\
  0 & 0 & r & 0 \\
  0 & 0 & 0 & r \sin\theta
  \end{array}
  \right)
  \ ,
\end{equation}
and the corresponding vierbein
$e_m{}^\mu$
is
\begin{equation}
\label{vierbein}
  e_m{}^\mu
  =
  \left(
  \begin{array}{cccc}
  \alpha & \beta_t & 0 & 0 \\
  0 & \beta_r & 0 & 0 \\
  0 & 0 & 1 / r & 0 \\
  0 & 0 & 0 & 1 / ( r \sin\theta )
  \end{array}
  \right)
  \ .
\end{equation}
The tetrad-frame connection coefficients
$\Gamma_{kmn}$ are
\begin{subequations}
\label{Gammaspherical}
\begin{align}
  \Gamma_{rtt}
  &=
  h_t
  \ ,
\\
  \Gamma_{rtr}
  &=
  h_r
  \ ,
\\
  \Gamma_{\theta t\theta}
  =
  \Gamma_{\phi t\phi}
  &=
  {\beta_t \over r}
  \ ,
\\
  \Gamma_{\theta r\theta}
  =
  \Gamma_{\phi r\phi}
  &=
  {\beta_r \over r}
  \ ,
\\
  \Gamma_{\phi\theta\phi}
  &=
  {\cot\theta \over r}
  \ ,
\end{align}
\end{subequations}
where $\beta_m$ are defined by equation~(\ref{betam}),
and $h_m$
are defined in terms of the vierbein coefficients by
\begin{subequations}
\label{hspherical}
\begin{align}
\label{htspherical}
  h_t
  &\equiv
  - \,
  \partial_r \ln\alpha
  \ ,
\\
\label{hrspherical}
  h_r
  &\equiv
  - \,
  \beta_t {\partial \ln\alpha \over \partial r}
  +
  {\partial \beta_t \over \partial r}
  -
  \partial_t \ln \beta_r
  \ .
\end{align}
\end{subequations}
A person at rest in the tetrad frame has, by definition,
tetrad-frame 4-velocity
$u^k = \{ 1 , 0 , 0 , 0 \}$,
and the covariant derivative $\DD_m u^r$
of the radial component $u^r$ of the tetrad-frame 4-velocity
of such a person is
\begin{equation}
\label{Dmur}
  \DD_m u^r
  =
  \Gamma^r_{tm}
  =
  h_m
  \ .
\end{equation}
Equations~(\ref{Dmur})
reveal that
$h_t$
and
$h_r$
represent
respectively
the proper radial acceleration
(minus the gravitational force),
and the ``Hubble parameter'' of the radial flow,
experienced by a person at rest in the tetrad frame.
Restricted to the $t$--$r$-plane,
$h_m \equiv \{ h_t , h_r \}$
constitute a tetrad-frame 2-vector.

The non-vanishing components of the tetrad-frame Riemann tensor
$R_{klmn}$
are
\begin{subequations}
\label{Riemannspherical}
\begin{align}
  R_{trtr}
  &=
  \DD_r h_t
  -
  \DD_t h_r
  \ ,
\\
  R_{t\theta t\theta}
  =
  R_{t\phi t\phi}
  &=
  - \,
  {1 \over r}
  \DD_t \beta_t
  \ ,
\\
  R_{t\theta r\theta}
  =
  R_{t\phi r\phi}
  &=
  - \,
  {1 \over r}
  \DD_t \beta_r
  =
  - \,
  {1 \over r}
  \DD_r \beta_t
  \ ,
\\
  R_{r\theta r\theta}
  =
  R_{r\phi r\phi}
  &=
  - \,
  {1 \over r} 
  \DD_r \beta_r
  \ ,
\\
  R_{\theta\phi \theta\phi}
  &=
  {2 M \over r^3}
  \ .
\end{align}
\end{subequations}
The non-vanishing components of the tetrad-frame Einstein tensor $G^{km}$ are
\begin{subequations}
\label{Gspherical}
\begin{align}
\label{Gttspherical}
  G^{tt}
  &=
  {2 \over r} 
  \left(
    - \, \DD_r \beta_r
    + {M \over r^2}
  \right)
  \ ,
\\
\label{Gtrspherical}
  G^{tr}
  &=
  {2 \over r}
  \DD_t \beta_r
  =
  {2 \over r}
  \DD_r \beta_t
  \ ,
\\
\label{Grrspherical}
  G^{rr}
  &=
  {2 \over r}
  \left(
    - \, \DD_t \beta_t - {M \over r^2}
  \right)
  \ ,
\\
\label{Gaaspherical}
  G^{\theta\theta}
  =
  G^{\phi\phi}
  &=
  D_r h_t - D_t h_r
  + 
  {1 \over r}
  \left(
  D_r \beta_r
  -
  D_t \beta_t
  \right)
  \ .
\end{align}
\end{subequations}
In spherically symmetric spacetimes,
the only
distinct non-vanishing component of the Weyl curvature tensor
is the scalar (spin-$0$) component,
the Weyl curvature scalar $C$
\begin{align}
  C
  &=
  \frac{1}{2}
  C_{trtr}
  =
  -
  \frac{1}{2}
  C_{\theta\phi\theta\phi}
  =
  -
  C_{t \theta t \theta}
  =
  -
  C_{t \phi t \phi}
\nn
  &=
  C_{r \theta r \theta}
  =
  C_{r \phi r \phi}
  =
  \frac{1}{6}
  (
  G^{tt}
  -
  G^{rr}
  +
  G^{\theta\theta}
  )
  -
  {M \over r^3}
  \ .
\end{align}
Thanks to the Bianchi identities,
the Einstein tensor automatically satisfies
covariant conservation
\begin{equation}
  D_m G^{mn}
  =
  0
  \ ,
\end{equation}
enforcing conservation of energy-momentum.
The two non-vanishing conservation equations
are those for energy and radial momentum,
which, written out in full, are
\begin{subequations}
\label{conserveG}
\begin{align}
  &
  \left(
  \partial_t
  +
  {2 \beta_t \over r}
  +
  h_r
  \right)
  G^{tt}
  +
  \left(
  \partial_r
  +
  {2 \beta_r \over r}
  +
  2 h_t
  \right)
  G^{tr}
\nn
  &
  \qquad\qquad
  + \,
  h_r \,
  G^{rr}
  +
  {2 \beta_t \over r} \,
  G^{\theta\theta}
  \  = \ 
  0
  \ ,
\\
  &
  \left(
  \partial_t
  +
  {2 \beta_t \over r}
  +
  2 h_r
  \right)
  G^{tr}
  +
  \left(
  \partial_r
  +
  {2 \beta_r \over r}
  +
  h_t
  \right)
  G^{rr}
\nn
  &
  \qquad\qquad
  + \,
  h_t \,
  G^{tt}
  -
  {2 \beta_r \over r} \,
  G^{\theta\theta}
  \  = \ 
  0
  \ .
\end{align}
\end{subequations}
Since each of the two conservation equations~(\ref{conserveG})
involves each of the 4 distinct non-vanishing components
$G^{tt}$, $G^{tr}$, $G^{rr}$, and $G^{\theta\theta}$
of the Einstein tensor,
it follows that any two of the Einstein equations
may be dropped in favor of the two equations
of energy-momentum conservation.

\section{Problem}
\label{problem}

This problem appeared on the final exam
of the first author's graduate course on general relativity
in Spring 2008.
Some of the equations below repeat equations elsewhere in this paper,
but they are left as is so that the problem remains self-contained.

\subsection*{The mechanism of mass inflation}

Einstein's equations in a spherically symmetric spacetime imply that
the covariant rate of change of the radial 4-gradient
$\beta_m \equiv \partial_m r = \{ \partial_t r , \partial_r r , 0 , 0 \}$
in the frame of any radially moving orthonormal tetrad
is
\begin{subequations}
\label{dtbsphericalapp}
\begin{align}
\label{dtbtsphericalapp}
  \DD_t \beta_t
  &=
  - \,
  {M \over r^2}
  -
  4 \pi r p
  \ ,
\\
\label{dtbrsphericalapp}
  \DD_t \beta_r
  &=
  4 \pi r f
  \ ,
\end{align}
\end{subequations}
where $\DD_t$ is the covariant time derivative,
$p$ is the radial pressure,
$f$ is the radial energy flux,
and $M$ is the interior mass defined by
\begin{equation}
\label{Mapp}
  {2 M \over r} - 1
  \equiv
  \beta^2
  \equiv
  - \beta_m \beta^m
  =
  {\beta_t^2} - {\beta_r}^2
  \ .
\end{equation}

\subsection{Freely-falling stream}

Consider a stream of matter
that is freely falling radially inside the horizon of
a spherically symmetric black hole.
Let $u$ be the radial component
of the tetrad-frame 4-velocity
$u^m$
of the stream relative to the ``no-going'' frame
where $\beta_r = 0$
(the frame of reference that divides ingoing frames
$\beta_r > 0$ from outgoing frames $\beta_r < 0$):
\begin{equation}
\label{um}
  u^m
  \equiv
  \{ - \beta_t / \beta , - \beta_r / \beta , 0 , 0 \}
  =
  \{ \sqrt{1 + u^2} , u , 0 , 0 \}
  \ .
\end{equation}
Note that $\beta_t$ is negative inside the horizon.
The time component
$u^t \equiv - \beta_t / \beta = \sqrt{ 1 + u^2 }$
of the 4-velocity
is positive (as it should be for a proper 4-velocity),
while the radial component
$u \equiv u^r \equiv - \beta_r / \beta$
of the 4-velocity
is negative ingoing, positive outgoing.
Show that along the worldline of the stream
\begin{subequations}
\label{dlnbudlnr}
\begin{align}
  {\dd \ln \beta \over \dd \ln r}
  &=
  {1 \over \beta^2}
  \left[
  - \,
  {M \over r}
  -
  4 \pi r^2
  \left(
  p + {\beta_r \over \beta_t}  f
  \right)
  \right]
  \ ,
\\
  {\dd \ln u \over \dd \ln r}
  &=
  {1 \over \beta^2}
  \left[
  {M \over r}
  +
  4 \pi r^2
  \left(
  p + {\beta_t \over \beta_r}  f
  \right)
  \right]
  \ .
\end{align}
\end{subequations}
[Hint:
If the stream is freely falling,
then the proper time derivative $\partial_t$ in the tetrad frame
of the stream equals the covariant time derivative $\DD_t$.
Thus the proper rates of change of $\ln \beta$ and $\ln u$
with respect to $\ln r$ along the worldline of the stream are
\begin{equation}
  {\dd \ln \beta \over \dd \ln r}
  =
  {\partial_t \ln \beta \over \partial_t \ln r}
  \ , \quad
  {\dd \ln u \over \dd \ln r}
  =
  {\partial_t \ln u \over \partial_t \ln r}
  \ .
\end{equation}
These can be evaluated through
\begin{subequations}
\begin{align}
\label{dtlnbeta}
  \partial_t \ln \beta
  &=
  \DD_t \ln \beta
  =
  {1 \over 2 \beta^2} \DD_t \beta^2
  =
  {1 \over 2 \beta^2} \DD_t ( \beta_t^2 - \beta_r^2 )
\nn
  &=
  {1 \over \beta^2} ( \beta_t \DD_t \beta_t - \beta_r \DD_t \beta_r )
  \ ,
\\
\label{dtlnu}
  \partial_t \ln u
  &=
  \DD_t \ln u
  =
  \DD_t \ln \beta_r - \DD_t \ln \beta
\nn
  &=
  {1 \over \beta_r}
  \DD_t \beta_r
  - \DD_t \ln \beta
  \ ,
\\
\label{dtlnr}
  \partial_t \ln r
  &=
  {1 \over r} \, \partial_t r
  =
  {\beta_t \over r}
  \ ,
\end{align}
\end{subequations}
with Einstein's equations~(\ref{dtbsphericalapp})
substituted into equations (\ref{dtlnbeta}) and (\ref{dtlnu}).]

\subsection{Equal ingoing and outgoing streams}

Consider the symmetrical case
of two equal streams of radially ingoing
($\beta_r > 0$)
and outgoing ($\beta_r < 0$)
neutral, pressureless, non-interacting matter (``dust''),
each of proper density $\rho$ in their own frames,
freely-falling into a charged black hole.
Show that
\begin{subequations}
\label{dlnbudlnrsym}
\begin{align}
\label{dlnbdlnrsym}
  {\dd \ln \beta \over \dd \ln r}
  &=
  - \,
  {1 \over 2 \beta^2}
  \left(
  - \,
  \lambda
  +
  \beta^2
  +
  \mu
  u^2
  \right)
  \ ,
\\
\label{dlnudlnrsym}
  {\dd \ln u \over \dd \ln r}
  &=
  - \,
  {1 \over 2 \beta^2}
  \left(
  \lambda
  -
  \beta^2
  +
  \mu
  +
  \mu
  u^2
  \right)
  \ ,
\end{align}
\end{subequations}
where
\begin{equation}
\label{lambdamu}
  \lambda
  \equiv
  Q^2/r^2 - 1
  \ , \quad
  \mu
  \equiv
  16 \pi r^2 \rho
  \ .
\end{equation}
Hence conclude that
\begin{equation}
\label{dlnbdlnusym}
  {\dd \ln \beta \over \dd \ln u}
  =
  {
  - \, \lambda + \beta^2 + \mu u^2
  \over
  \lambda - \beta^2 + \mu + \mu u^2
  }
  \ .
\end{equation}
[Hint:
The assumption that the streams are neutral, pressureless, and non-interacting
is needed to make the streams freely-falling,
so that equations~(\ref{dlnbudlnr}) are valid.
The pressure $p$ in the tetrad frame of each stream is
the sum of the electromagnetic pressure $p_e$ and the streaming pressure $p_s$
\begin{equation}
  p = p_e + p_s
  \ .
\end{equation}
The electromagnetic pressure $p_e$ is
\begin{equation}
\label{pe}
  p_e
  =
  - \, {Q^2 \over 8 \pi r^4}
  \ ,
\end{equation}
with $Q$ the charge of the black hole,
which is constant because the infalling streams are neutral.
The streaming pressure $p_s$ that each stream sees is
\begin{equation}
  p_s
  =
  \rho ( u_s^r )^2
  \ ,
\end{equation}
where the streaming 4-velocity $u_s^m$
between the two streams is
the 4-velocity of the observed stream Lorentz-boosted
by the 4-velocity of the observing stream
(the radial velocities $u^r$ of the observed and observing streams
have opposite signs)
\begin{subequations}
\begin{align}
  u_s^t
  &=
  ( u^t )^2 + ( u^r )^2 = 1 + 2 u^2
  \ ,
\\
  u_s^r
  &=
  - 2 u^t u^r
  =
  - 2 u \sqrt{ 1 + u^2 }
  \ .
\end{align}
\end{subequations}
The energy flux $f$ in the tetrad frame of each stream is
the streaming flux $f_s$
\begin{equation}
  f
  =
  f_s
  =
  \rho u_s^t u_s^r
  \ .
\end{equation}
\begin{subequations}
You should find that the combinations of streaming pressure and flux that
go into equations~(\ref{dlnbudlnr}) are
\begin{align}
  p_s + {\beta_r \over \beta_t} f_s
  &=
  2 \rho u^2
  \ ,
\\
  p_s + {\beta_t \over \beta_r} f_s
  &=
  - 2 \rho ( 1 + u^2 )
  \ .
\end{align}
\end{subequations}
]

\subsection{Reissner-Nordstr\"om phase}

If the accretion rate is small,
then initially the stream density $\rho$ is small,
and consequently $\mu$ is small.
Argue that in this regime
equation~(\ref{dlnbdlnusym}) simplifies to
\begin{equation}
\label{dlnbdlnurn}
  {\dd \ln \beta \over \dd \ln u}
  =
  {
  - \, \lambda + \beta^2
  \over
  \lambda - \beta^2
  }
  \ .
\end{equation}
Hence conclude that
\begin{equation}
\label{betarn}
  \beta =
  {C \over u}
  \ ,
\end{equation}
where $C$ is some constant set by initial conditions
(generically, $C$ will be of order unity).

\subsection{Transition to mass inflation}

Argue that in the Reissner-Nordstr\"om phase, $\beta$ becomes small,
and $u$ grows large,
as the streams fall to smaller radius $r$.
Argue that in due course equation~(\ref{dlnbdlnusym})
becomes well-approximated by
\begin{equation}
\label{dlnbdlnumi}
  {\dd \ln \beta \over \dd \ln u}
  =
  {
  - \, \lambda + \mu u^2
  \over
  \lambda + \mu u^2
  }
  \ .
\end{equation}
Treating $\lambda$ and $\mu$ as constants (which is a good approximation),
show that the solution to equation~(\ref{dlnbdlnumi}) subject to
the initial condition set by equation~(\ref{betarn}) is
\begin{equation}
\label{betami}
  \beta
  =
  {C ( \lambda + \mu u^2 ) \over \lambda u}
  \ .
\end{equation}
[Hint:
$\lambda$ is positive.
In the Reissner-Nordstr\"om solution,
$\beta$ would go to zero at the inner horizon.]

\subsection{Sketch}

Sketch the solution~(\ref{betami}),
plotting $u$ against $\beta$ on logarithmic axes.
Mark the regime where mass inflation is occurring.

\subsection{Inflationary growth rate}

Argue that during mass inflation the inflationary growth rate
$\dd \ln \beta / \dd \ln r$
is
\begin{equation}
  {\dd \ln \beta \over \dd \ln r}
  =
  -
  {\lambda^2
  \over
  2 C^2 \mu}
  \ .
\end{equation}
Comment on
how the inflationary growth rate depends on accretion rate (on $\rho$).

\section{Multiple streams}
\label{multiple}

Let there be multiple neutral, pressureless, radially freely-falling streams.
Analogously to equation~(\ref{um}),
let $u_a$ denote the radial component of the 4-velocity
$u_a^m$
of stream $a$ relative to the no-going frame:
\begin{equation}
  u_a^m
  \equiv
  \{ - \beta_{a,t} / \beta , - \beta_{a,r} / \beta , 0 , 0 \}
  =
  \{ \sqrt{1 + u_a^2} , u_a , 0 , 0 \}
  \ .
\end{equation}
The velocity $u_a$ is negative ingoing, positive outgoing.
Let $u_{ba}^m$ denote the 4-velocity of stream $b$
relative to stream $a$
\begin{equation}
\label{uba}
  u_{ba}^t
  =
  u_b^t u_a^t
  -
  u_b^r u_a^r
  \ , \quad
  u_{ba}^r
  =
  u_b^r u_a^t
  -
  u_b^t u_a^r
  \ .
\end{equation}
The proper pressure seen by stream $a$
is the sum of the electromagnetic pressure $p_e$,
equation~(\ref{pe}),
which is the same for every stream,
and the streaming pressure $p_a$,
which is a sum over the contributions from all other streams $b$:
\begin{equation}
  p_a = \sum_b \rho_b ( u_{ba}^r )^2
  \ .
\end{equation}
The energy flux $f_a$ seen by stream $a$
equals the streaming energy flux,
which is again a sum over the contributions from all other streams $b$:
\begin{equation}
  f_a = \sum_b \rho_b u_{ba}^t u_{ba}^r
  \ .
\end{equation}
Equations~(\ref{dlnbudlnr})
remain valid individually for each freely-falling stream.
The combinations of streaming pressure and flux
that go into equations~(\ref{dlnbudlnr}) are
\begin{subequations}
\label{pfs}
\begin{align}
\label{ps}
  p_a + {\beta_{a, r} \over \beta_{a, t}} f_a
  &=
  \sum_b
  \rho_b ( u_b^r )^2
  \left(
  1 - {u_a^r u_b^t \over u_a^t u_b^r}
  \right)
  \ ,
\\
\label{fs}
  p_a + {\beta_{a, t} \over \beta_{a, r}} f_a
  &=
  -
  \sum_b
  \rho_b ( u_b^t )^2
  \left(
  1 - {u_a^t u_b^r \over u_a^r u_b^t}
  \right)
  \ .
\end{align}
\end{subequations}
In the particular case of two streams,
one ingoing and one outgoing,
that are streaming relativistically through each other,
as is the relevant case during mass inflation,
equations~(\ref{pfs}) simplify to
\begin{subequations}
\label{pftwo}
\begin{align}
  p_a + {\beta_{a, r} \over \beta_{a, t}} f_a
  &\approx
  2 \rho_b u_b^2
  \ ,
\\
  p_a + {\beta_{a, t} \over \beta_{a, r}} f_a
  &\approx
  -
  2 \rho_b u_b^2
  \ .
\end{align}
\end{subequations}
Here $a$ represents either of the ingoing or outgoing streams,
and $b$ represents the other stream.
Inserting expressions~(\ref{pfs}) into equations~(\ref{dlnbudlnr}) yields
\begin{subequations}
\label{dlnbudlnrtwo}
\begin{align}
\label{dlnbdlnrtwo}
  {\dd \ln \beta \over \dd \ln r_a}
  &=
  - \,
  {1 \over 2 \beta^2}
  \left(
  - \,
  \lambda
  +
  \beta^2
  +
  \mu_b
  u_b^2
  \right)
  \ ,
\\
\label{dlnudlnrtwo}
  {\dd \ln u_a \over \dd \ln r_a}
  &=
  - \,
  {1 \over 2 \beta^2}
  \left(
  \lambda
  -
  \beta^2
  +
  \mu_b
  u_b^2
  \right)
  \ ,
\end{align}
\end{subequations}
where $\lambda$ is the same as before, equation~(\ref{lambda}),
and the ``accretion rate''
$\mu_b$
of each stream is given by
equation~(\ref{mub}),
generalizing equation~(\ref{mu}).
Note that it is important to distinguish the radii $r_a$ and $r_b$
used as the independent variable on the left hand sides of
equations~(\ref{dlnbudlnrtwo}),
since the two streams follow different trajectories into the black hole.
Equation~(\ref{dlnbdlnrtwo}) divided by equation~(\ref{dlnudlnrtwo})
yields equation~(\ref{dlnbdlnutwo}) in the text.

It is essential to
the derivation of equations~(\ref{pftwo}) and (\ref{dlnbudlnrtwo})
from (\ref{pfs})
that one of the streams be ingoing and the other outgoing.
If there are two streams that are both ingoing or both outgoing,
and if both streams are relativistic
relative to the no-going frame
(as will be the case on the approach to the inner horizon),
then instead of the two expressions~(\ref{pftwo})
being approximately equal and of opposite sign,
the two expressions are approximately equal and of the same sign.
The ratio
of the two equations~(\ref{dlnbudlnrtwo}) is then
$\dd \ln \beta / \dd \ln u_a = - 1$,
and mass inflation is not ignited.


\section{Unexpected values of parameters}
\label{unexpected}

It was commented at the end of \S\ref{inflationtwo}
that the parameters
$\lambda$, $\mu_b$, and $\mu_d$
of the analytic fit~(\ref{betaami})
to the model illustrated in Figure~\ref{bu}
are perhaps surprising given the boundary conditions of the model.
The parameters are listed in Table~\ref{par}.

One surprising feature is that
the charge-to-mass ratio $\Qbh/\Mbh = 0.8$
of the black hole set by the boundary conditions at the sonic point
would predict $\lambda = 3$ at the inner horizon
of the corresponding Reissner-Nordstr\"om black hole,
smaller than the value $\lambda = 14.6$
actually measured in the model near the inner horizon.
The reason for this is that
the baryons are charged, and therefore repelled by the black hole,
so the effective charge-to-mass of the black hole near the inner horizon
is smaller than $\Qbh/\Mbh = 0.8$,
and $\lambda$ is consequently larger.

A second surprising feature of the measured parameters
is that the ``accretion rate'' parameter
$\mu_d = 1.39 \times 10^{-2}$ of the dark matter
exceeds that $\mu_b = 1.93 \times 10^{-3}$ of the baryons,
in spite of the fact that the boundary conditions at the sonic point
set the dark matter density to be much less than the baryonic density,
$\rho_d / \rho_b = 10^{-3}$.
Ultimately,
the reason for this is again that the baryons are charged not neutral.
Although the dark-matter-to-baryon ratio
$\rho_d / \rho_b = 10^{-3}$
at the sonic point may seem small,
in fact it is almost as large as can be
given the large charge-to-mass ratio $\Qbh/\Mbh = 0.8$ of the black hole,
and the small value
$w = 10^{-6}$
of the baryonic equation of state.
If the ratio of neutral dark matter to baryons at the sonic point
is increased,
then the black hole is inclined to become more neutral,
and it becomes impossible to achieve the desired charge-to-mass ratio
$\Qbh/\Mbh = 0.8$
(as discussed by \cite{Hamilton:2004av},
the mathematical condition that the parameters must satisfy
is that the radial 4-gradient $\beta_m$ must be spacelike
at the sonic point outside the outer horizon).
A small
$w = 10^{-6}$
makes it harder to achieve a large
charge-to-mass ratio
$\Qbh/\Mbh = 0.8$,
because then the sonic point is far from the black hole,
and there is more room for the charged baryons to be repelled
by the black hole as opposed to falling in.
Thus one should think of
the dark-matter-to-baryon ratio
$\rho_d / \rho_b = 10^{-3}$
of the model as being ``large'' rather than small.
This is why the dark matter ``accretion rate'' $\mu_d$
measured near the inner horizon
exceeds the baryonic rate $\mu_b$,
Table~\ref{par}.
If the dark-matter-to-baryon ratio at the sonic point were set
to a truly small value, then indeed the dark matter $\mu_d$
would be less than the baryonic $\mu_b$ near the inner horizon.

\section{Mass accretion rate}
\label{massaccretionrate}

This paper parameterizes the accretion rate
of the black hole as the rate
$\Mbhdot$
of increase of its mass $\Mbh$
as measured by distant observers,
whereas \cite{Hamilton:2004av,Hamilton:2004aw}
parameterized the rate by a quantity
$\eta_s \equiv \alpha r / t$
evaluated at the sonic point,
where the boundary conditions are set.
What is the relation between the two?

As in \cite{Hamilton:2004av},
the charge $\Qbh$
and
mass $\Mbh$
of the black hole at any instant
are defined to be those that would be measured by a distant observer
if there were no charge or mass outside the sonic point,
\begin{equation}
  \Qbh
  =
  Q
  \mbox{ and }
  \Mbh
  =
  M + {Q^2 \over 2 r}
  \mbox{ at the sonic point}
  \ .
\end{equation}
The extra mass $Q^2 / 2 r$ added to the interior mass $M$
is the mass-energy in the electric field outside the sonic point,
given no charge outside the sonic point.

In self-similar solutions,
the black hole mass increases linearly with time,
$\Mbh \propto t$,
and the mass accretion rate $\Mbhdot$ is therefore
\begin{equation}
\label{Mbhdot}
  \Mbhdot
  \equiv
  {\dd \Mbh / \dd t}
  =
  {\Mbh / t}
  \ ,
\end{equation}
where $t$ is time measured at rest at infinity.

The time $t$ measured at infinity coincides with the
proper time recorded on dust (dark matter) clocks
that free-fall radially from rest at infinity.
This can be seen as follows.
First,
for the line element~(\ref{metric}),
an interval of proper time recorded on a clock at rest in the tetrad frame
is $\dd t / \alpha$.
This is true because
a person at rest in the tetrad frame has, by definition,
tetrad-frame 4-velocity $u^m = \{ 1 , 0 , 0 , 0 \}$,
so their coordinate-frame 4-velocity is
$u^\mu = e_m{}^\mu u^m = e_t{}^\mu = \{ \alpha , \beta_t , 0 , 0 \}$,
that is,
$\dd r / \dd t = \beta_t / \alpha$ and
$\dd \theta / \dd t = \dd \phi / \dd t = 0$,
and it then follows from the line element~(\ref{metric})
that the proper time interval of a person at rest in the tetrad frame
is $\dd t / \alpha$ as asserted
(the line element is in fact constructed to have this property).
Second,
for a freely-falling tetrad,
the vierbein coefficient
$\alpha$ is a function only of $t$, not $r$.
This follows from equation~(\ref{htspherical}) and the fact that, by definition,
the proper acceleration $h_t$ experienced in a freely-falling frame vanishes.
The fact that $\alpha(t)$ is a function only of $t$,
not of $r$, expresses coordinate gauge freedom in the choice of time $t$.
It is natural to fix the gauge
by setting the coordinate time $t$ equal to the proper time
at rest at infinity.
This is equivalent to setting $\alpha = 1$
on dust clocks that free-fall from rest at infinity.
In other words,
as claimed,
time $t$ measured at rest at infinity coincides with the
proper time recorded on dust clocks
that free-fall radially from rest at infinity.
To avoid any confusion,
what this means is that if a dust clock falling through
the sonic point says the time is $t$, and another dust clock falling
through the sonic point a little later says the time is $t + dt$,
then an observer at rest at infinity will say that the two instants
are separated by time $\dd t$.
One might worry about the difference in light travel times
between sonic point and observer at the two instants,
but for realistically small accretion rates
the difference is negligible.

In order to translate this time at infinity into what is happening
at the sonic point, it is necessary to know the transformation
between the dust frame and the baryonic frame at the sonic point,
which is equivalent to knowing the vierbein coefficients $\beta_m$ of each.
The coefficients for the baryonic frame are set by the
boundary conditions at the sonic point.
The interior mass at the sonic point, a gauge-invariant scalar, fixes
$\beta_t^2 - \beta_r^2$ in all frames.
The only possible uncertainty is in the $\beta_r$ of the dust
at the sonic point.
Technically, the dust $\beta_r$ depends on the radial profile of mass-energy
through which the dust has fallen.
However, if the mass-energy outside the sonic point is neglected,
an excellent approximation for realistically low accretion rates,
then $\beta_r = 1$ at the sonic point,
which is the value adopted here and in \cite{Hamilton:2004av,Hamilton:2004aw}.

By assumption,
self-similar solutions possess conformal time translation invariance,
which is to say that they remain invariant
under a scale transformation of $t$ at fixed $r / t$.
It follows that $\xi^m$ defined by
\begin{equation}
  r \xi^m \partial_m
  =
  \left. {\partial \over \partial \ln t} \right\rvert_{r/t}
  =
  t {\partial \over \partial t}
  +
  r {\partial \over \partial r}
  =
  ( t e^m{}_t + r e^m{}_r ) \partial_m
\end{equation}
is a dimensionless conformal Killing vector, or homothetic vector.
Comparison to the inverse vierbein~(\ref{inversevierbein})
shows that the time component of the dimensionless homothetic vector is
$\xi^t = t e^t{}_t / r = t / ( \alpha r )$.
As argued above,
the coefficient $\alpha$ equals one
in the particular case of dust that free-falls from rest at infinity.
Thus the time $t$ at rest at infinity is
\begin{equation}
  t
  =
  r \xi_d^t
  \ ,
\end{equation}
where $\xi_d^t$ is the time component of the dimensionless homothetic vector
in the dust frame.
Thus finally the mass accretion rate~(\ref{Mbhdot}) is
\begin{equation}
  \Mbhdot =
  {\Mbh \over r \xi_d^t}
  \quad
  \mbox{at the sonic point}
  \ .
\end{equation}
By comparison,
the accretion rate parameter $\eta_s$ of \cite{Hamilton:2004av,Hamilton:2004aw} was
\begin{equation}
  \eta_s
  =
  {1 \over \xi^t}
  \quad
  \mbox{at the sonic point}
  \ ,
\end{equation}
where $\xi^t$ is the time component of the dimensionless homothetic vector
in the baryonic frame.

\end{document}